\documentclass[journal]{IEEEtran}

\usepackage{xcolor}
\usepackage[colorlinks]{hyperref}
\usepackage[nolist]{acronym}
\usepackage[normalem]{ulem}
\usepackage{booktabs}
\usepackage{gensymb}
\usepackage{tikz}
\usepackage{animate}
\usepackage{array}	
\usepackage{multirow}
\usetikzlibrary{arrows}

\newcommand{\tabfigure}[2]{\raisebox{-.5\height}{\includegraphics[#1]{#2}}}

\ifCLASSOPTIONcompsoc
\usepackage[caption=false,font=normalsize,labelfont=sf,textfont=sf]{subfig}
\else
\usepackage[caption=false,font=footnotesize]{subfig}
\fi

\hypersetup{
	colorlinks=true, 
	linkcolor=blue!70!black, 
	citecolor=red!70!black, 
	filecolor=blue!70!black, 
	urlcolor=blue!70!black, 
}

\newacro{MIMO}[MIMO]{multiple-input and multiple-output}
\newacro{FEM}[FEM]{finite element method}

\newcommand\figwidth{8.9} 

\usepackage{graphicx}
\graphicspath{{figures/}}

\usepackage{capekCommands}

\newacro{MIMO}[MIMO]{\mbox{multiple-input} \mbox{multiple-output}}
\newacro{TARC}[TARC]{total active reflection coefficient}
\newacro{RMS}[RMS]{root mean square}

\usepackage[normalem]{ulem}

\newcommand{\RWQ}[1]{\texttt{#1}}
\newcommand{\RWA}[1]{{#1}\vspace{0.5cm}}
\renewcommand{\RS}[1]{\textcolor{red}{\ifmmode\text{\sout{\ensuremath{#1}}}\else\sout{#1}\fi}}

\newif\ifarxiv
\arxivfalse 
\arxivtrue  

\begin{document}
\ifarxiv\else
\newpage
\thispagestyle{empty}
\onecolumn

\renewcommand{\thefigure}{\Alph{figure}}
\renewcommand{\theequation}{A.\arabic{equation}}

\begin{center}
	\textbf{Submission questions}
\end{center}

\section{What is the problem being addressed by the manuscript and why is it important to the Antennas and Propagation community? (limited to 100 words)}

A long lasting, yet not fully resolved, problem of uncorrelated channels' existence and their selective excitation is rigorously solved in this paper through the application of the point group theory. It is demonstrated that the key property in this task is geometry symmetry possessed by a radiating body. As far as some symmetries are present, a fixed number of current schemes can be excited in such a way that the antenna metrics, \eg{}, radiation patterns, are decoupled. These findings and method for their utilization are of interest for all multi-port and MIMO antenna theorist and designers.

\section{What is the novelty of your work over the existing work? (limited to 100 words).}

The presence of symmetries is connected with the possibility of decoupling current schemes on an antenna. The maximum number of these schemes and the minimum number of feeding points needed to excite them is rigorously determined from the point group theory. The role of ports placed at the symmetry planes is discussed. A general framework to deal with symmetries is presented, including a solution to the combinatorial optimization problem of feeding synthesis (determination of ports’ placement, amplitude and phase). Such a complete treatment cannot be found in existing works.

\section{Provide up to three references, published or under review, (journal papers, conference papers, technical reports, etc.) done by the authors/coauthors that are closest to the present work. Upload them as supporting documents if they are under review or not available in the public domain. Enter N.A. if it is not applicable.}

\begin{itemize}
    \item M. Masek, M. Capek, L. Jelinek, and K. Schab, ``Modal Tracking Based on Group Theory," {\it IEEE Trans. Antennas Propag.\/}, vol. 68, no. 2, pp. 927-–937, Feb. 2020.
    \item M. Capek, L. Jelinek, and M. Masek, ``Optimality of total active reflection coefficient and realized gain for multi-port lossy antennas," submitted to {\it IEEE Trans. Antennas Propag.\/}
    \item M. Masek, M. Capek, and L. Jelinek, ``Feeding Positions Providing the Lowest TARC of~Uncorrelated Channels," accepted to {\it 2020 14th European Conference on Antennas and Propagation (EuCAP)}, Copenhagen, Denmark, 2020.
\end{itemize}

\section{Provide up to three references (journal papers, conference papers, technical reports, etc.) done by other authors that are most important to the present work. Enter “N.A.” if it is not applicable.}

\begin{itemize}
    \item J. Andersen and H. Rasmussen, ``Decoupling and descattering networks for antennas," {\it IEEE Trans. Antennas Propag.\/}, vol. 24, no. 6, pp. 841–-846, Nov. 1976.
    \item N. Peitzmeier and D. Manteuffel, ``Upper bounds and design guidelines for realizing uncorrelated ports on multimode antennas based on symmetry analysis of characteristic modes," {\it IEEE Trans. Antennas Propag.\/}, vol. 67, no. 6, pp. 3902–-3914, June 2019.
    \item J.-M. Hannula, T. O. Saarinen, J. Holopainen, and V. Viikari, ``Frequency reconfigurable multiband handset antenna based on a multichannel transceiver," {\it IEEE Trans. Antennas Propag.\/}, vol. 65, no. 9, pp. 4452–-4460, Sept. 2017.
\end{itemize}

\newpage
\setcounter{section}{0}
\begin{center}
	\textbf{Response letter}
\end{center}
Dear Editor,
have a nice day!

\section*{Reviewer I}
\RWQ{...}

\RWA{...}

\section*{Reviewer II}
\RWQ{...}

\RWA{...}

\section*{Reviewer III}
\RWQ{...}

\RWA{...}

\clearpage
\setcounter{section}{0}
\setcounter{page}{1}
\setcounter{equation}{0}
\setcounter{figure}{0}
\renewcommand{\thefigure}{\arabic{figure}}
\renewcommand{\theequation}{\arabic{equation}}
\newpage
\twocolumn
\fi


\title{Excitation of Orthogonal Radiation States}
\author{Michal~Masek, Lukas~Jelinek, and Miloslav~Capek, \IEEEmembership{Senior Member, IEEE}
\thanks{Manuscript received  \today; revised \today. This work was supported by the Czech Science Foundation under project~\mbox{No.~19-06049S} and by the Grant Agency of the Czech Technical University in Prague under grant \mbox{SGS19/168/OHK3/3T/13}.}
\thanks{M. Masek, L. Jelinek, and M. Capek are with the Department of Electro-magnetic Field, Czech Technical University in Prague, Prague 166 27, Czech Republic (e-mails:   michal.masek@fel.cvut.cz; lukas.jelinek@fel.cvut.cz; miloslav.capek@fel.cvut.cz).}
\ifarxiv\else%
\thanks{Color versions of one or more of the figures in this communication are
available online at http://ieeexplore.ieee.org.}
\thanks{Digital Object Identifier XXX}
\fi%
}    
\ifarxiv\else%
\markboth{Journal of \LaTeX\ Class Files,~Vol.~XX, No.~XX, \today}{Masek \MakeLowercase{\textit{et al.}}: Modal Tracking Based on Group Theory}
\fi%

\maketitle

\begin{abstract}
A technique of designing antenna excitation realizing orthogonal states is presented. It is shown that a symmetric antenna geometry is required in order to achieve orthogonality with respect to all physical quantities. A maximal number of achievable orthogonal states and a minimal number of ports required to excite them are rigorously determined from the knowledge of an antenna's symmetries. The number of states and number of ports are summarized for commonly used point groups (a rectangle, a square, etc.). The theory is applied to an example of a rectangular rim where the positions of ports providing the best total active reflection coefficient, an important metric in multi-port systems, are determined. The described technique can easily be implemented in existing solvers based on integral equations.
\end{abstract}

\begin{IEEEkeywords}
Antenna theory, computer simulation, eigenvalues and eigenfunctions, electromagnetic modeling, method of moments, modal analysis.
\end{IEEEkeywords}

\section{Introduction}
\label{sec:intro}

\IEEEPARstart{T}{he} ever-growing requirements on data throughput capacity \cite{GiordaniEtAl_Towards6GNetworks} and simultaneous full occupancy of the radio spectrum has led to many novel concepts in recent decades \cite{Reed_NewTechnologiesSolvingSpectrumShortage_2016}. One of the most successful techniques is the \ac{MIMO} method~\cite{GesbertEtAl_overviewMIMO, Winter_CapacityOfRadioCommunationSystems} heavily utilized in modern communication devices~\cite{JensenWallace_ReviewForMIMOWirelessCommunications, YangHanzo_FiftyYearsOfMIMO}. When considering \ac{MIMO} spatial multiplexing, spatial correlation has a strong impact on ergodic channel capacity~\cite{2010_Molisch_Book}, therefore, low mutual coupling between the states generated by individual antennas is required~\cite{AndersenRasmussen_DecouplingAndSescatteringNetworksForAntennas, Kildal_CorrelationAndCapacityOfMIMOSystems}. 

In this paper, free-space channel capacity is increased by considering spatial multiplexing realized by orthogonal electromagnetic field states excited by a multi-port radiator~\cite{DaviuFabresGalloBataller_DesignOfAMultimodeMIMOantennaUsingCM, Saarinen_etal_CombinatoryFeedingMethodForMobileApplication, Hanulla_etal_FrequencyReconfiguragleMultibandHandset}. This assumes that orthogonal states are a good starting position for weakly correlated realistic channels where stochastical effects cannot be neglected. Instead of an array of transmitters~\cite{CoetzeeYu_PortDecouplingForSmallArraysByMeansOfAnEigenmodeFeedNetwork}, the orthogonality is provided by a general multiport antenna system. This approach addresses the question of how many orthogonal states can, in principle, be induced by a radiating system of a given geometry and how many localized ports are needed to excite them separately. 

Previous research on this topic utilized characteristic modes \cite{HarringtonMautz_TheoryOfCharacteristicModesForConductingBodies, HarringtonMautz_PatternSynthesisForLoadedNportScatterers} which provide orthogonal states in far field. Unfortunately, as shown by the long history of attempts within the characteristic mode community~\cite{ PeitzmeierManteuffel_SelectiveExcitationCMs_EuCAP,  Ethier_2009_TCM_MIMO_decoupled,  ChaudhurySchroederChaloupka_MIMOAntennaBasedOnOrthogonalityCMs,  LiShi_PatternRecinfigurableMIMOUsingCMs,  SuEtAl_RadiatonEnergyAndMutualCouplingForMIMOAntennaCMs, LiMiersLau_DesignOfMIMOhandsetAntennasBasedOnCMmanipulationAtFrequencyBandsBelow1GHz, Manteuffel_Martens-CompactMultimodeMultielementAntennaForIndoorUWB, JaafarCollardeySharaiha_OptimizedmanipulationOfNetworkCMsForWidebandSmallAntennaMatching, Wenetal_DesignOfMIMOAntennaForSmartwatchUsingTCM, QuetAl_MIMOAntennasUsingControlledOrthogonalCMs, WuSuLiSu_ReductionInOutOfBandAntennaCouplingUsingCMs}, this task is nearly impossible to accomplish, as entire-domain functions defined over arbitrarily shaped bodies cannot be selectively excited by discrete ports~\cite{Ethier_2009_TCM_MIMO}.

Many other methods exist to characterize and approach the maximal capacity, for both a special case of spherical geometry~\cite{GustafssonNordebo_CharacterizationOfMIMOAntennasUsingSVW,GustafssonNordebo_SpectralEfficiencyOfSphere} and for arbitrarily shaped antennas~\cite{XimenesAlmeida_CapacityOfMIMOSystemsUsingSWE, GlazunovEtAl_PhysicalLimitationsOfInteractionOfSphereAndRandomField,EhrenborgGustafsson_FundamentalBoundsOnMIMOAntennas}. The number of degrees of freedom represented by electromagnetic field states were studied on an information theory level as well~\cite{Migliore_RoleOfNumberOfDOFinMIMOChannels, GlazunovZhang2011_OptimalMIMOantennaCoefs, Migliore_HorseIrMoreImporatntThanHorseman}. As with the characteristic modes approach, in all these cases the optimal coefficients do not prescribe any particular excitation of a selected or optimized antenna designs. This issue was solved in~\cite{Mohajer2010_SWEforMIMOsystems} utilizing a singular value decomposition of excitation coefficients represented in spherical wave expansion and in~\cite{Capeketal_OptimalityOfTARCAndRealizedGainForMultiPortAntennas_Arxiv} by employing a port-mode basis~\cite{Mautz1973}. The orthogonal radiation patterns are excited, however, the schemes are not orthogonal with respect to other physical operators, leading to unpleasant effects, such as non-zero mutual reactances~\cite{WallaceJensen_MutualCouplingInMIMOWirelessSystems}.

The situation changes dramatically for a structure invariant under certain symmetry operations, including rotation, reflection, or inversion. Certain symmetry operations were utilized in \cite{DaviuFabresGalloBataller_DesignOfAMultimodeMIMOantennaUsingCM} and \cite{KrewskiSchroederSolbach_Multiband2portMIMOLTEantennaDesign}, however, a general approach can be reached only by applying point group theory \cite{McWeeny_GroupTheory} which allows modes computed by arbitrary modal decomposition to be classified into several irreducible representations (irreps) which are orthogonal to each other. Spherical harmonics~\cite{Hamermesh_GroupTheoryandItsApplicationToPhysicalProblems_1989} of a different order are a notable example of such an uncorrelated set of states. A known property of physical states selected arbitrarily from two different irreps is that all mutual metrics are identically zero \cite{SchabEtAl_EigenvalueCrossingAvoidanceInCM}. This useful property has already been utilized for the block-diagonalization of the bodies of a revolution matrix \cite{Knorr_1973_TCM_symmetry} and further study reveals interesting properties regarding the simultaneous excitation of perfectly isolated states \cite{MartensManteuffel_SystematicDesignMethodOfMobileMultipleAntennaSystemUsingCM,YangAdams_SystematicShapeOptimizationOfSymmetricMIMOAntennasUsingCM}. An additional benefit is that selective excitation is possible since the antenna excitation vectors may follow the irreducible representations of the underlying structure \cite{Krewski_2011_TCM_MIMO}.

The key instrument employed in this work is the group theory-based construction of a symmetry-adapted basis \cite{McWeeny_GroupTheory} and block-diagonalization of the operators. This methodology leads to a fully automated design, without the necessity of a visual inspection or manual manipulation of the data \cite{Peitzmeier_Manteuffel-UpperBoundsForUncorrelatedPorts2019}. The upper bound on the number of orthogonal states and the lower bound on the number of ports are rigorously derived only from the knowledge of symmetries. It is observed that the later number is significantly lower than the number of ports utilized in practice~\cite{PeitzmeierManteuffel_SelectiveExcitationCMs_EuCAP}. The placement of a given number of ports maximizing a selected antenna metric is investigated through combinatorial optimization~\cite{PapadimitriouSteiglitz_CombinatorialOptimization} over vector adapted bases.

The entire design procedure can easily be incorporated into a simple algorithm, thus opening possibilities to analyze \ac{MIMO} antennas automatically. All findings are demonstrated on a set of canonical geometries. The figure of merit classifying the performance of \ac{MIMO} radiating systems is the \ac{TARC} \cite{2005_ManteghiRahmatSamii_TARC}, however, all the presented material is general and valid for all operators and all metrics. 

The paper is structured as follows. The theory is developed in Section~\ref{sec:orthogonalChannels}, primarily based on point group theory and eigenvalue decomposition. The basic consequences are demonstrated on an example in Section~\ref{sec:illustrativeExample}. Section~\ref{sec:ExcitationStatesPG} addresses the important questions of how many orthogonal states are available and how many ports are needed to excite them independently. The optimal placement of a given number of ports is then solved in Section~\ref{sec:PortPositioning} via an exhaustive search. The paper is concluded in Section~\ref{sec:conclusion}.


\section{Orthogonal Channels}
\label{sec:orthogonalChannels}

Let us assume antenna metric~$p$ defined via quadratic form
\begin{equation}
    p\left(\V{q}_m, \V{q}_n\right) = \left\langle \V{q}_m, \OP{A}\left(\V{q}_n\right) \right\rangle,
\end{equation}
where $\V{q}_m$ and $\V{q}_n$ are states of the system (\eg{}, modal current densities, modal far fields, or excitation states, see Table~\ref{tab:introTable}), $\OP{A}$ is a linear complex operator, see~Appendix~\ref{sec:ap-operators} for representative examples, and $\langle\cdot, \cdot\rangle$ denotes the inner product
\begin{equation}
    \langle \V{a}\left(\V{r}\right), \V{b}\left(\V{r}\right) \rangle = \int_\varOmega \V{a}^*(\V{r})\cdot\V{b}(\V{r})\D{V},
    \label{eq:innerProduct}
\end{equation}
where $\V{a}\left(\V{r}\right)$ and $\V{b}\left(\V{r}\right)$ are generic vector fields supported in region~$\varOmega$, $\V{r} \in \varOmega$. For the purpose of this paper, orthogonality of states is further defined as
\begin{equation}
    p\left(\V{q}_m \in \OP{S}_i, \V{q}_n \in \OP{S}_j\right) =  \zeta_{ijmn}\delta_{ij},
    \label{eq:orthogonality}
\end{equation}
where $\left\lbrace\OP{S}_i\right\rbrace$ are disjoint sets of states~$\V{q}$, $\zeta$ are normalization constants and $\delta_{ij}$ is a Kronecker delta.

\begin{table}[]
    \centering
    \caption{Three examples of system states~$\V{q}_m$ and associated operators~$\OP{A}$ preserving orthogonality in the sense of~\eqref{eq:orthogonality}. The algebraic representation of states $\M{V}_m$, and operators $\M{A}$, is expressed in a basis~$\{\basisFcn_n (\V{r})\}$, see Appendix~\ref{sec:ap-operators} for details. All quantities depicted in the table are subsequently introduced throughout the paper.}
    \begin{tabular}{ccc}
    current densities & far fields & excitation \\
    characteristic modes~\cite{HarringtonMautz_TheoryOfCharacteristicModesForConductingBodies} & far-field patterns~\cite{GustafssonTayliEhrenborgEtAl_AntennaCurrentOptimizationUsingMatlabAndCVX} & port modes~\cite{1978_Harrington_TAP} \\ \toprule
    \tabfigure{scale=0.85}{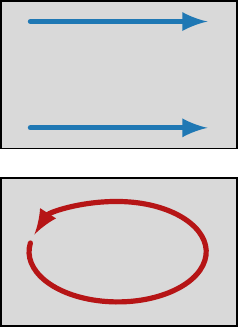} &
    \tabfigure{scale=0.875}{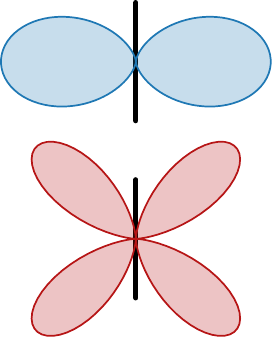} &
    \tabfigure{}{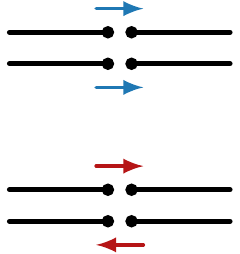} \\ \midrule
    $\V{q}_m = \V{J}_m (\V{r})$ & $\V{q}_m = \V{F}_m (\vartheta, \varphi)$ & $\V{q}_m = \V{E}^\T{i}_m (\V{r})$ \\
    $\OP{A}_1 = \OP{X}_0$, $\OP{A}_2 = \OP{R}_0$ & $\OP{A} = \OP{R}_0 = \RE\left\{\OP{Z}_0\right\}$ & $\OP{A} = y = z^{-1}$ \\ \midrule
    $\M{Q}_m = \Ivec_m$ & $\M{Q}_m = \Ivec_m$ & $\M{Q}_m = \M{v}_m$ \\
    $\M{A}_1 = \M{X}_0$, $\M{A}_2 = \M{R}_0$ & $\M{A} = \M{R}_0 = \RE\left\{ \M{Z}_0 \right\}$ & $\M{A} = \M{P}^\herm \M{Z}^{-1} \M{P}$ \\ \bottomrule
    \end{tabular}
    \label{tab:introTable}
\end{table}

In order to obtain a numerically tractable problem, procedures such as the \ac{MoM}~\cite{Harrington_FieldComputationByMoM} or \ac{FEM}~\cite{Volakis_1998_FEM} are commonly employed, recasting states~$\V{q}$, operators~$\OP{A}$, and sets $\OP{S}$ into column vectors~$\M{Q}$, matrices~$\M{A}$~\cite{Shilov_LinearAlgebra}, and linear vector spaces $S$, respectively, see Table~\ref{tab:introTable} and Appendix~\ref{sec:ap-operators}. Within such a paradigm, the orthogonality~\eqref{eq:orthogonality} can be written as
\begin{equation}
    \M{Q}_m^\herm \M{A} \M{Q}_n = 0: \quad 
    \M{Q}_m \in S_i, \quad \M{Q}_n \in S_j,
    \label{eq:orthogonalityMatrix}
\end{equation}
which means that matrix $\M{A}$ is block-diagonalized in the basis generated by these states.

Difficulties in finding orthogonal sets of vectors strongly depend on the number of operators $\left\{\M{A}_i \right\}$ with respect to which relation \eqref{eq:orthogonalityMatrix} must simultaneously be satisfied. In the case of a sole operator $\left\{\M{A} \right\}$ or two operators $\left\{\M{A}_1, \M{A}_2 \right\}$, the solution to a standard $\M{A} \M{Q} = \lambda \M{Q}$ or a generalized $\M{A}_1 \M{Q} = \lambda \M{A}_2 \M{Q}$ eigenvalue problem gives vectors which diagonalize the underlying operators~\cite{Wilkinson_AlgebraicEigenvalueProblem}. The well-known example is the characteristic modes decomposition~\cite{HarringtonMautz_TheoryOfCharacteristicModesForConductingBodies} defined as
\begin{equation}
    \Xmat_0 \Ivec_m = \lambda_m \Rmat_0 \Ivec_m,
    \label{eq:CMs}
\end{equation}
where $\Ivec_m$ are the characteristic modes, $\lambda_m$ are the characteristic numbers, and $\Zmat_0 = \Rmat_0 + \J \Xmat_0$ is the vacuum impedance matrix defined in Appendix~\ref{sec:ap-operators}. Multiplying~\eqref{eq:CMs} from the left by the $n$th characteristic mode~$\Ivec_n$ and considering unitary radiated power of each mode, we see that matrices~$\M{X}_0$ and $\M{R}_0$ are diagonalized,
\begin{align}
    \dfrac{1}{2}\Ivec_n^\herm \Xmat_0 \Ivec_m & = \lambda_n \delta_{mn}, \\
    \dfrac{1}{2}\Ivec_n^\herm \Rmat_0 \Ivec_m & = \delta_{mn},
\end{align}
generating orthogonality in reactive and radiated power, respectively. In the case of three or more operators, simultaneous diagonalization is possible only under special conditions (\eg{}, mutually commuting matrices). For example, choosing a third matrix~$\Wmat = \omega \partial \Xmat_0 / \partial \omega$, \cite{CapekJelinek_OptimalCompositionOfModalCurrentsQ}, it is realized that
\begin{equation}
\dfrac{1}{2}\Ivec_n^\herm \Wmat \Ivec_m = w_{mn} \neq w_{mn} \delta_{mn},
\label{eq:energyModes}
\end{equation}
\ie{}, characteristic modes, in general, only diagonalize matrices~$\Xmat_0$ and~$\Rmat_0$. However, when point symmetries are present, at least simultaneous block-diagonalization can be reached and, as explained in the following sections, orthogonal states with respect to all operators describing the physical behaviour of the underlying structure can be easily established.

\subsection{Orthogonal States Based on Point Symmetries}
\label{sec:orthogonalChannels:PS}

\begin{figure}[t]
    \centering
    \includegraphics[width = 8.9cm]{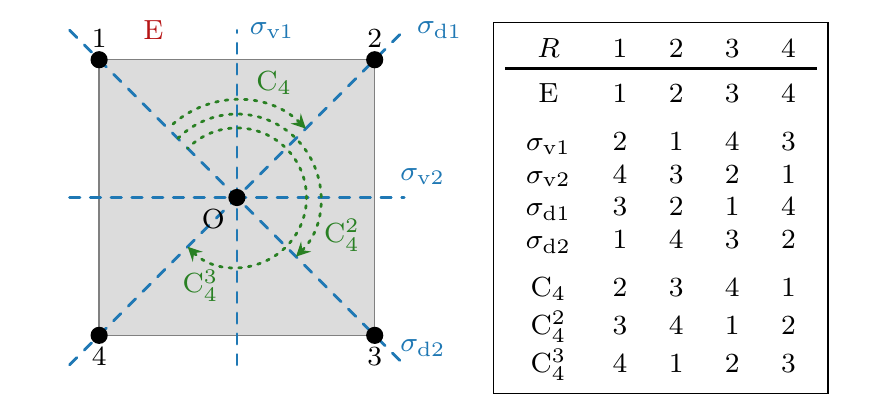}\\
    \caption{An example of symmetry operations -- a square. This structure belongs to point group $\T{C}_{4\T{v}}$~\cite{McWeeny_GroupTheory} and has eight symmetry operations: identity $\T{E}$, four reflections $\sigma$ and three rotations $\T{C}$. The table shows how each node is transformed via each symmetry operation.}
    \label{fig:symmetriesExamples}
\end{figure}

\begin{figure}[t]
    \centering
    \includegraphics{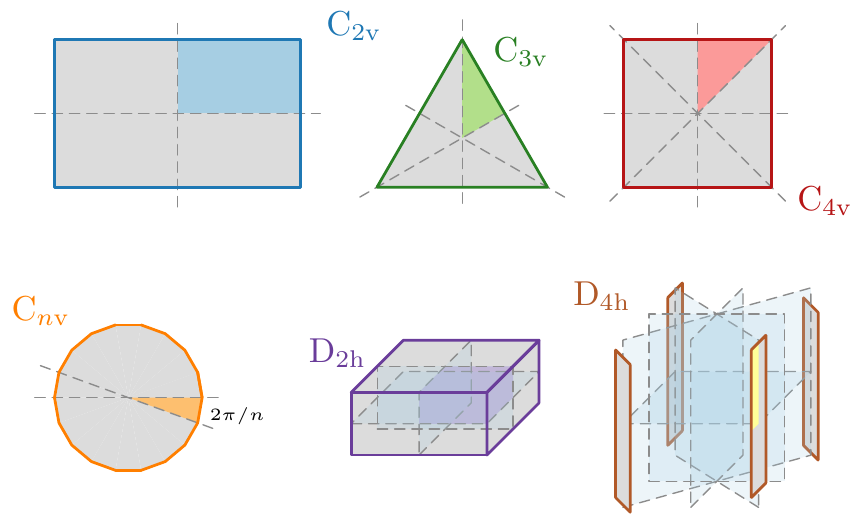}
    \caption{Examples of various point groups. Colored regions show generator of the structure, \ie{}, the minimal part of the object from which the entire structure can be constructed via symmetry operations.}
    \label{fig:groupsExamples}
\end{figure}

In the case of symmetrical objects (see examples of symmetry operations in Fig.~\ref{fig:symmetriesExamples} and sketches of several point groups in Fig.~\ref{fig:groupsExamples}), point group theory~\cite{McWeeny_GroupTheory} shows that physical states of the system can be uniquely divided into disjoint sets called species. For each such set, a rectangular matrix $\M{\Gamma}^{(\alpha, i)}$ can be constructed so that
\begin{equation}
    \widehat{\M{A}}^{(\alpha, i)} = \left(\M{\Gamma}^{(\alpha, i)}\right)^\trans \M{A} \M{\Gamma}^{(\alpha, i)}
    \label{eq:blockDiagonalization}
\end{equation}
is a single block of a block-diagonalized matrix $\widehat{\M{A}} = \M{\Gamma}^\trans \M{A} \M{\Gamma}$ with matrix~$\M{\Gamma}$ accumulating all blocks $\M{\Gamma}^{(\alpha, i)}$ side by side. Indices $\alpha$ and $i$ form species $(\alpha, i)$, with $\alpha$ denoting selected irreducible representation (irrep) and $i\left(\alpha\right) \in \{1, \dots , g^{(\alpha)}\}$ counting along a dimension of the selected irrep~\cite{McWeeny_GroupTheory}. The rectangular matrix $\M{\Gamma}^{(\alpha, i)}$ will be called a symmetry-adapted basis and its construction within the \ac{MoM} paradigm is detailed in~\cite{Maseketal_ModalTrackingBasedOnGroupTheory}.

Let us recall once again the characteristic modes~\eqref{eq:CMs} and their lack of orthogonality with respect to stored energy~\eqref{eq:energyModes}. Possessing a symmetrical structure and considering block-diagonalization~\eqref{eq:blockDiagonalization} of matrices~$\Rmat_0$, $\Xmat_0$, and~$\Wmat$, the characteristic modes
\begin{equation}
    \Ivec_m^{(\alpha,i)} = \M{\Gamma}^{(\alpha,i)} \widehat{\Ivec}_m^{(\alpha,i)}
    \label{eq:CMreduced1}
\end{equation}
with
\begin{equation}
    \widehat{\Xmat}_0^{(\alpha,i)} \widehat{\Ivec}_m^{(\alpha,i)} = \lambda_m \widehat{\Ivec}_m^{(\alpha,i)} \widehat{\Rmat}_0^{(\alpha,i)},
\end{equation}
belong exclusively to species~$\left(\alpha,i\right)$~\cite{Maseketal_ModalTrackingBasedOnGroupTheory}. In such case, the relation~\eqref{eq:energyModes} changes to
\begin{equation}
\dfrac{1}{2}\left(\Ivec_n^{(\alpha,i)}\right)^\herm \Wmat \Ivec_m^{(\beta,j)} = w_{mn} \delta_{\alpha\beta} \delta_{ij},
\end{equation}
which means that the characteristic modes from different irreps ($\alpha \neq \beta$), or from the same irrep but different dimension ($i \neq j$), are orthogonal with respect to stored energy as well. This statement can be generalized to all operators resulting from a \ac{MoM} paradigm, see examples in Appendix~\ref{sec:ap-operators} or in~\cite{JelinekCapek_OptimalCurrentsOnArbitrarilyShapedSurfaces}.

The relation~\eqref{eq:blockDiagonalization} states that columns of matrices $\M{\Gamma}^{(\alpha, i)}$ form vector spaces~$S$ in~\eqref{eq:orthogonalityMatrix}, consequently the columns of $\M{\Gamma}^{(\alpha, i)}$ can be desired vectors~$\M{Q}$. In such a case, the orthogonality~\eqref{eq:orthogonalityMatrix} holds simultaneously for all operators $\left\lbrace\M{A}_i\right\rbrace$ describing the physical behavior of the underlying structure whenever two vectors belong to different species.

\section{Illustrative example}
\label{sec:illustrativeExample}
This section demonstrates the usefulness of the point group-based block diagonalization~\eqref{eq:blockDiagonalization} to obtain orthogonal states.

\begin{figure}[t]
    \centering
	\includegraphics{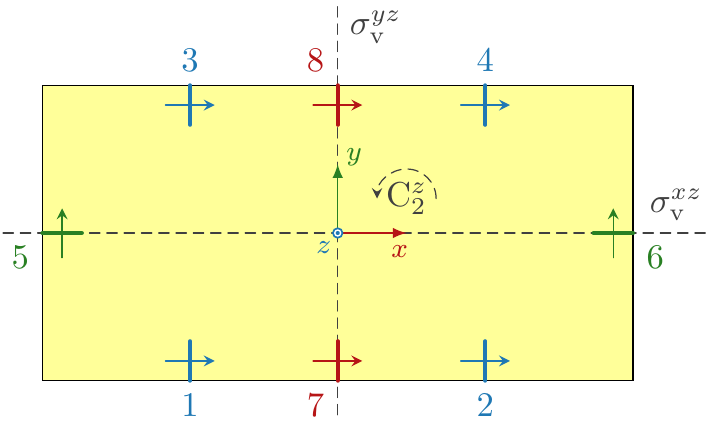}
	\caption{Port locations on a rectangular plate. Reprinted from \cite[Fig.~5]{PeitzmeierManteuffel_SelectiveExcitationCMs_EuCAP}. Arrows show the orientation of the voltage sources.}
	\label{fig:dirkPlate}
\end{figure}

The design procedure is illustrated on the example of a rectangular plate of dimensions $2L \times L$ and of electrical size $ka \approx 10.19$ ($k$ abbreviates a free-space wavenumber and $a$ denotes the radius of the smallest sphere circumscribing the plate), which was used in~\cite{PeitzmeierManteuffel_SelectiveExcitationCMs_EuCAP} to construct orthogonal states via the selective excitation of~\acp{CM}. The~\acp{CM} in~\cite{PeitzmeierManteuffel_SelectiveExcitationCMs_EuCAP} were visually separated into four \Quot{groups} (using the nomenclature of~\cite{PeitzmeierManteuffel_SelectiveExcitationCMs_EuCAP}), and voltage sources (ports) were associated with each such group so as to provide maximum excitation of the dominant \ac{CM} of each group. In order to independently control four sets of modes, eight voltage sources (delta gaps) were used. The structure and positions of voltage sources used in~\cite{PeitzmeierManteuffel_SelectiveExcitationCMs_EuCAP} are shown in Fig.~\ref{fig:dirkPlate}. Unit voltages were considered with polarity determined by the second column of Table~\ref{tab:dirkPlateVoltage}.

The point group theoretical treatment introduced in Section~\ref{sec:orthogonalChannels:PS} offers a different solution to the same problem. The underlying object has four point symmetries (identity, rotation of $\pi$ around $z$-axis and two reflections via $xz$ and $yz$ planes) and belongs to the $\T{C}_{2\T{v}}$ point group (see Fig.~\ref{fig:groupsExamples}) which possess four one-dimensional irreps~\cite{McWeeny_GroupTheory}. The number of distinct species\footnote{Only one-dimensional irreps exist in this case, \ie{}, dimensionality of each irrep $\alpha$ is $g^{(\alpha)} = 1$.} introduced in Section~\ref{sec:orthogonalChannels:PS} is four, each being connected to a distinct matrix $\M{\Gamma}^{(\alpha, 1)}$. Within a standard notation~\cite{McWeeny_GroupTheory}, these irreps are listed in the third column of Table~\ref{tab:dirkPlateVoltage}.

As mentioned in Section~\ref{sec:orthogonalChannels:PS}, any columns of matrices~$\M{\Gamma}^{(\alpha, 1)}$~\cite{Maseketal_ModalTrackingBasedOnGroupTheory} can be used as excitation vectors $\M{V}^{(\alpha, 1)}$, see Appendix~\ref{sec:ap-excitationVector}, to enforce orthogonality. To minimize the number of voltage sources used, it is advantageous to select those columns which have non-zero elements at the same positions across all species. In the specific case of~Fig.~\ref{fig:dirkPlate}, matrices $\M{\Gamma}^{(\alpha, 1)}$ also contain columns with only four non-zero entries (\ie{}, with four voltage sources) at positions corresponding to ports $1$--$4$ shown in Fig.~\ref{fig:dirkPlate} in blue. Orientations of connected unit voltage sources are shown in the last column of Table~\ref{tab:dirkPlateVoltage}. This means that the eight ports used in~\cite{PeitzmeierManteuffel_SelectiveExcitationCMs_EuCAP} are not necessary to provide four orthogonal states.

This example introduces a series of questions of fundamental importance for multiport and multimode devices:
\begin{enumerate}
    \item[Q1)] How many orthogonal states, with respect to all physical operators $N_\T{s}$, can be found for a structure belonging to a specific point group?
    \item[Q2)] What is the lowest number of ports $N_\T{p}$ that ensures a given number of orthogonal states?
    \item[Q3)] Where should ports be placed to maximize the performance of a device, with respect to a given physical metric, to maintain the orthogonality of states?
\end{enumerate}

These questions are addressed throughout the paper using point group theory revealing important aspects of the symmetry-based design of orthogonal states.

\begin{table}[t] 
    \centering 
    \caption{Orthogonal excitation states for a plate from Fig.~\ref{fig:dirkPlate}. The second column is a solution found in \cite[Table~IV]{PeitzmeierManteuffel_SelectiveExcitationCMs_EuCAP}. Numbers refer to the voltage source in Fig.~\ref{fig:dirkPlate} and superscripts $+$ and $-$ denote its orientation with respect to the directions in Fig.~\ref{fig:dirkPlate}. The third column denotes irreps designation in the notation of point group theory~\cite{McWeeny_GroupTheory}. The last column shows a solution via the only four voltage sources described in this paper.}
    \begin{tabular}{cccc}
        Set & Ports \cite[Table~IV]{PeitzmeierManteuffel_SelectiveExcitationCMs_EuCAP} & irrep $\alpha$ & Four ports \\ \toprule
        $\OP{S}_1$ & $1^+, 2^-, 3^+, 4^-$ & $\T{A}_1$ & $1^+, 2^-, 3^+, 4^-$\\
        $\OP{S}_2$ & $5^+, 6^+$ & $\T{B}_1$ & $1^+, 2^-, 3^-, 4^+$\\ 
        $\OP{S}_3$ & $7^+, 8^+$ & $\T{B}_2$ & $1^+, 2^+, 3^+, 4^+$\\
        $\OP{S}_4$ & $7^+, 8^-$ & $\T{A}_2$ & $1^+, 2^+, 3^-, 4^-$\\
         \bottomrule
    \end{tabular} 
    \label{tab:dirkPlateVoltage}
\end{table}

\section{Excitation States Based on Point Group Theory}
\label{sec:ExcitationStatesPG}
Referring to~\cite[eq.~(16)]{Maseketal_ModalTrackingBasedOnGroupTheory}, symmetry-adapted excitation vectors can be constructed as
\begin{equation}
  	\M{V}^{(\alpha, i)}\left(\xi\right) = \frac{g^{(\alpha)}}{g} \sum_{R \in G} \contragradient{d}^{(\alpha)}_{ii} \left(R \right) \M{C} \left(R \right) \M{V}\left(\xi\right),
  	\label{eq:symmetryAdaptedVec}
\end{equation}
which is a linear map from excitation vector~$\M{V}\left(\xi\right) \in \mathbb{C}^{N_\T{u}\times 1}$ (see Appendix~\ref{sec:ap-excitationVector}) onto a symmetry-adapted excitation vector~$\M{V}^{(\alpha, i)}\left(\xi\right) \in \mathbb{C}^{N_\T{u}\times 1}$ that satisfies
\begin{equation}
\left(\M{V}^{(\alpha,i)}(\xi)\right)^\herm \M{A} \M{V}^{(\beta,j)}(\xi) = \zeta_{\alpha\beta ij} \delta_{\alpha\beta} \delta_{ij}
\label{eq:orthogonalitySAV}
\end{equation}
for an arbitrary operator~$\M{A}\in \mathbb{C}^{N_\T{u}\times N_\T{u}}$ with $N_\T{u}$ being number of unknowns (number of basis functions). The mapping~\eqref{eq:symmetryAdaptedVec} is characterized by the point group of structure $G = \left\lbrace R \right\rbrace$ consisting of symmetry operations~$R$, dimensionality~$g^{(\alpha)} = \T{dim} \, \M{D}^{(\alpha)}$ of irrep $\alpha$, the order of the point group~\mbox{$g = \sum_\alpha \left(g^{(\alpha)}\right)^2$}, mapping matrix $\M{C}\left(R\right)$ and irreducible matrix representation \mbox{$\M{D}^{(\alpha)} = \left[d_{ij}^{(\alpha)}\right]$} with $\mbox{$\contragradient{\M{D}} = \left(\M{D}^{-1}\right)^\T{T}$}$, see~\cite{McWeeny_GroupTheory} and \cite[Sec. II-C]{Maseketal_ModalTrackingBasedOnGroupTheory} for more details. The application of~\eqref{eq:symmetryAdaptedVec} and the exact meaning of all variables used is illustrated in an example in Appendix~\ref{sec:ap-symmetryAdaptedVector}.

Throughout the paper, excitation vector~$\M{V}\left(\xi\right)$ represents an arbitrarily shaped port (\eg{}, delta-gap, coaxial probe, etc.) that lies entirely in the generator of the structure, see highlighted areas in~Fig.~\ref{fig:groupsExamples}, and variable~$\xi$ is used to code the position of this port. As an example, assume that port No.~1 in Fig.~\ref{fig:dirkPlate} is a delta-gap port represented by excitation vector~$\M{V}\left(1\right)$. Notice that it is placed in one of the quadrants, which are the generators of the structure. Each summand of~\eqref{eq:symmetryAdaptedVec} maps (changing orientation, position and amplitude) this port on its symmetry positions 2, 3, 4, creating symmetry-adapted excitation vector~$\M{V}^{(\alpha, i)}\left(1\right)$ for a particular species~$(\alpha, i)$.

The first two questions from Section~\ref{sec:illustrativeExample} can be answered by inspecting~\eqref{eq:symmetryAdaptedVec}: 
\begin{enumerate}
    \item The maximum number of orthogonal states,~$N_\T{s}$, (orthogonal with respect to all physical operators) is equal to the number of species of the given point group, \ie{}, to the number of vectors~$\M{V}^{(\alpha, i)}\left(\xi\right)$ generated by~\eqref{eq:symmetryAdaptedVec} for a given set of ports in the generator of the structure, which is $N_\T{s} = \sum_\alpha g^{(\alpha)}$. In other words, for a given distribution of ports in the generator of the structure, described by vector $\M{V}\left(\xi\right)$, there exist $N_\T{s}$ ways of how to symmetry-adapt this vector within the given point group. Each symmetry-adaptation creates an orthogonal excitation vector $\M{V}^{\left(\alpha, i\right)} \left(\xi\right)$.
    \item The minimum number of ports, $N_\T{p}$, needed to distinguish all orthogonal states mentioned above is equal to the number of symmetry operations in point group $G$ since each summand of~\eqref{eq:symmetryAdaptedVec} maps initial excitation vector~$\M{V} \left(\xi\right)$  onto a new position and there are as many summands as symmetry operations. See detailed example in Appendix~\ref{sec:ap-symmetryAdaptedVector}. It is assumed that each mapping is unique, otherwise not all orthogonal states are reached -- this possibility is discussed later in Section~\ref{sec:ExcitationStatesPG:badPlacing}.
\end{enumerate}
Table~\ref{tab:excitationEfficiency} summarizes the number of maximal reachable orthogonal states and number of ports required for it for the known point symmetry groups.

When combined together, the answers to Q1 and Q2 show how orthogonal states can be efficiently established for a given point group. On the other hand, this procedure does not ensure that all states lead to the same/optimal value of the selected antenna metric. This calls for a reply to question~Q3 which is addressed in Section~\ref{sec:PortPositioning}. 

\begin{table}[t]
    \caption{Maximum number of symmetry-based orthogonal states~$N_\T{s}$~/ minimal number of ports~$N_\T{p}$ needed to excite all of them for a given point group. Selected point groups are shown in Fig.~\ref{fig:groupsExamples}. A Schoenflies notation~\cite{McWeeny_GroupTheory} is used for point groups naming.}
    \label{tab:excitationEfficiency}
    \includegraphics[width = \figwidth cm]{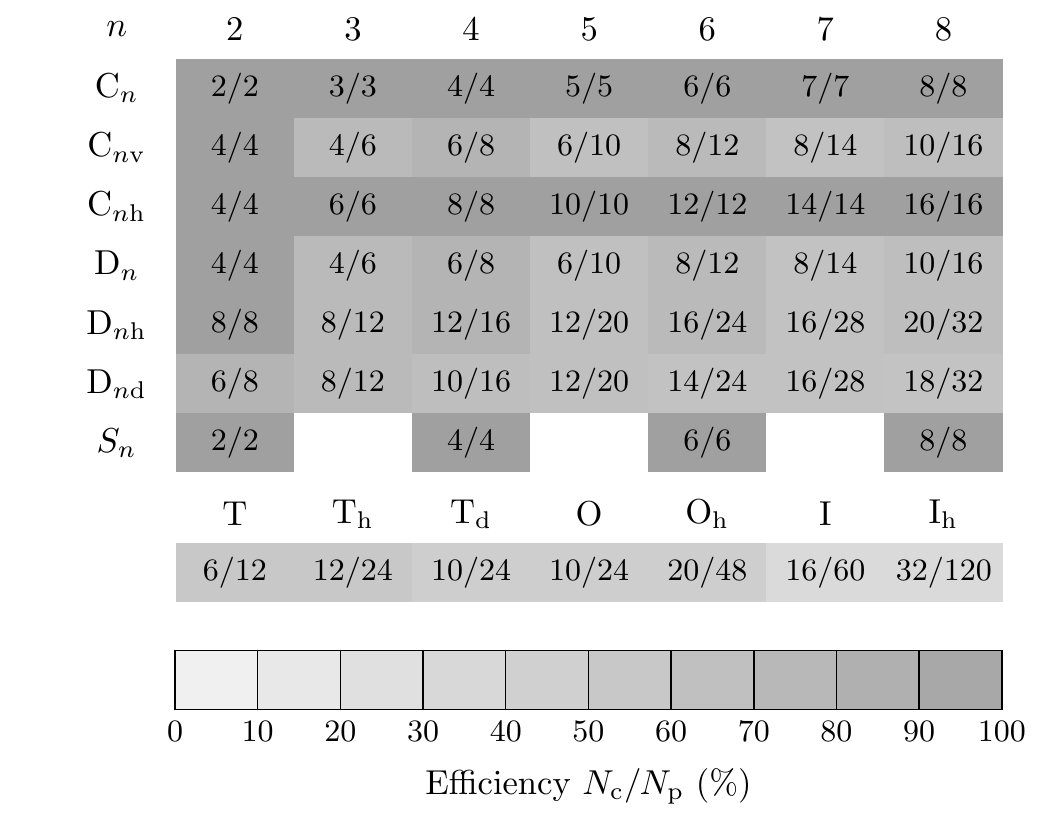}
\end{table}

\subsection{Port Placed in the Reflection Plane}
\label{sec:ExcitationStatesPG:badPlacing}
Formula \eqref{eq:symmetryAdaptedVec} suggests that a problematic design appears when the port corresponding to excitation vector~$\M{V}\left(\xi\right)$ lies at the boundary of the generator of the structure~\cite{McWeeny_GroupTheory}, \eg{}, at the reflection plane. In this case, the port generally breaks the symmetry of the structure making the process of symmetry adaptation invalid. To give an example, imagine that a delta-gap port is placed at position~$\xi = 5$ in Fig.~\ref{fig:dirkPlate}. The reflection~$\sigma^{xz}_\T{v}$ and identity operation~$\T{E}$ project this port onto itself but with different polarity. This collision is demonstrated in Table~\ref{tab:placementCollision}. In this case, only states belonging to irreps $\T{A}_2$ and $\T{B}_2$ are realizable. More than four ports would be needed to establish four states.

\begin{table}[t] 
    \centering 
    \caption{A symmetry-adapted delta gap number five from Fig.\ref{fig:dirkPlate}.}
    \begin{tabular}{ccccc}
        $R~\backslash~\alpha$ & $\T{A}_1$ & $\T{A}_2$ & $\T{B}_1$ & $\T{B}_2$ \\ \toprule
        $\T{E}$       & $5^+$ & $5^+$ & $5^+$ & $5^+$\\
        $\sigma^{xz}_\T{v}$ & $5^-$ & $5^+$ & $5^-$ & $5^+$\\ 
        $\sigma^{yz}_\T{v}$ & $6^+$ & $6^-$ & $6^-$ & $6^+$\\
        $\T{C}_2^z$     & $6^-$ & $6^-$ & $6^+$ & $6^+$\\
         \bottomrule
    \end{tabular} 
    \label{tab:placementCollision}
\end{table}

\section{Ports' Positioning}
\label{sec:PortPositioning}

\begin{figure}[t]
    \centering
    \includegraphics[width = 0.5\textwidth]{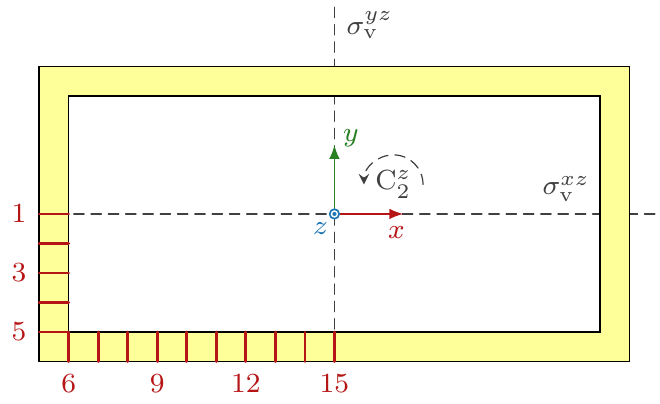}
    \caption{The structure of the rectangular rim. Possible placements of ports~$\xi$ in the generator of the structure are highlighted in red.}
    \label{fig:rectangularRim:mesh}
\end{figure}

In order to answer the third question from Section~\ref{sec:illustrativeExample} -- \textit{Where should ports be placed to maximize the performance of a device, with respect to a given physical metric, to maintain the orthogonality of states?} -- it is necessary to take into account the particular requirements on the performance of the device. An example of investigating port positions to optimize the \ac{TARC} of an antenna is used to demonstrate the sequence of steps to resolve this question. Instead of the rectangular plate shown in Fig.~\ref{fig:dirkPlate}, a rectangular rim of dimensions $2 L \times L$ and width $L/10$ is considered, see the object in Fig.~\ref{fig:rectangularRim:mesh}. The geometry of the rim belongs to the same point group as the plate but allows for the placement of discrete ports~\cite{Balanis1989} at an arbitrary position without creating undesired short circuits.

\subsection{Total Active Reflection Coefficient}
The \acl{TARC}~\cite{2005_ManteghiRahmatSamii_TARC}, which is defined as 
\begin{equation}
    \label{eq:TARC:def}
    t = \sqrt{1-\frac{P_\T{rad}}{P_\T{in}}},
\end{equation}
is used as a performance metric, where $P_\T{rad}$ stands for radiated power and $P_\T{in}$ stands for incident power. Within the \ac{MoM} framework, \eqref{eq:TARC:def} can be reformulated as
\begin{equation}
    \label{eq:TARC:Vp}
    t\left(\M{v}\right) = \sqrt{1 - \frac{4 Z_0 \M{v}^\herm{} \M{P}^\herm{} \M{Y}^\herm{} \M{R}_0 \M{Y} \M{P} \M{v}}{ \M{v}^\herm{} \M{k}^\herm{}\M{k} \M{v} }},
\end{equation}
where
\begin{equation}
    \label{eq:Amat}
    \M{k} = \M{e} + Z_0\M{y}.
\end{equation}
Here,~$\M{e}$ is the identity matrix, \mbox{$Z_0 = 50\;\Omega$} is the characteristic impedance of all transmission lines connected to the ports, $\M{Y} = \M{Z}^{-1} \in \mathbb{C}^{N_\T{u}\times N_\T{u}}$ is an admittance matrix, $\M{R}_0$ is the radiation part of the impedance matrix, and $\M{y}\in\mathbb{C}^{N_\T{p} \times N_\T{p}}$ is the admittance matrix seen by $N_\T{p}$ connected ports. Each port is represented by one column of matrix~$\M{P}$ and port voltages are all accumulated in vector~$\M{v}$. Matrix~$\M{P}$ is therefore of size~${N_\T{u} \times N_\T{p}}$ and the excitation vector is given by~$\M{V} = \M{P} \M{v}$, see Appendixes~\ref{sec:ap-operators}, \ref{sec:ap-excitationVector} and \ref{sec:ap-tarc} for detailed derivations.

\subsection{Optimization Problem}
\label{sec:optimProblem}
The problem of \ac{TARC} minimization with additional constraints on $N_\T{m}$~orthogonal states is defined as to find port excitation vectors~$\left\{\M{v}_m\right\}$, $m \in \left\{1,\dots,N_\T{m}\right\}$ and port configuration~$\M{P}$ such as to fulfill
\begin{equation}
    \begin{array}{cccc}
        \underset{\left\lbrace\M{P},\M{v}_m\right\rbrace}{\T{minimize}} \quad & t_{\T{RMS}} & & \\
        \T{subject\, to} \quad & \M{v}_m^\herm \M{P}^\herm \M{A}_1 \M{P} \M{v}_n & = & \,\, 0,\quad m \neq n, \\
        & \quad \quad \vdots \quad \quad & = & \,\,\, \vdots \\
        & \M{v}_m^\herm \M{P}^\herm \M{A}_{N_\T{k}} \M{P} \M{v}_n & = & \,\, 0,\quad m \neq n,
    \end{array}
    \label{eq:optimProblem}
\end{equation}
where the \ac{RMS} metric based on~\eqref{eq:TARC:Vp} is adopted
\begin{equation}
    t_{\T{RMS}} = \sqrt{\frac{1}{N_\T{m} N_\T{f}} \sum\limits_{m=1}^{N_\T{m}} \sum\limits_{f=1}^{N_\T{f}} t^2\left(\M{v}_m, \omega_f \right)}
    \label{eq:RMS_freq}
\end{equation}
to measure the overall performance over~$N_\T{f}$ frequency samples~$\omega_f$ and over multiple states. Matrices~$\M{A}_k$, $k \in \left\{1,\dots,N_\T{k}\right\}$, in the constraints above are placeholders for matrix operators from, \eg{}, Appendix~\ref{sec:ap-operators}. These constraints enforce simultaneous orthogonality with respect to all operators describing the physical system at hand, \eg{}, with respect to far fields ($\M{A}_k = \Ymat^\herm \M{R}_0 \Ymat$), current densities ($\M{A}_k = \Ymat^\herm \Ymat$), excitation vectors ($\M{A}_k$ is the identity matrix), or energy stored by the states ($\M{A}_k = \Ymat^\herm \M{W} \Ymat$).

In light of the discussion in Section~\ref{sec:orthogonalChannels}, the simultaneous realization of all $N_\T{k} > 2$ constraints in~\eqref{eq:optimProblem} is only possible on symmetric structures and only when excitation vectors~$\Vvec_m = \M{P} \M{v}_m$ are given by~\eqref{eq:symmetryAdaptedVec}, \ie{}, $\M{V}_m = \M{V}^{(\alpha, i)}=\M{P}\M{v}^{(\alpha, i)}$. This imposes specific requirements on port matrix~$\M{P}$ and port voltages~$\M{v}^{(\alpha, i)}$.

First, ports represented by columns of port matrix $\M{P}$ have to be symmetrically distributed on the structure. This is achieved by placing a port (a single column of matrix~$\M{P}$) at arbitrary position~$\xi$ in the generator of the structure and then by replication of this port by the application of symmetry operations~$R \in G$ (column of matrix~$P$ is transformed by mapping matrices $\M{C}(R)$). Each replication results in a new port, \ie{}, new column\footnote{Note, that one of the symmetry operations is an identity which represents the original port in the generator of the structure and the corresponding column of matrix~$\M{P}$ should be omitted.} of port matrix $\M{P}$.

Second, port excitation vector $\M{v}$ is constructed so that only ports placed in the region of the generator of the structure are excited (others are kept at zero voltage) and the symmetry-adaptation~\eqref{eq:symmetryAdaptedVec} of vector $\M{V}(\xi) = \M{P}\M{v}$ is processed. Here and further, $\xi$~represents a particular position in the generator of the structure, see possible placements in Fig.~\ref{fig:rectangularRim:mesh}.

Lastly, port voltages $\M{v}^{(\alpha, i)}$ for species ${(\alpha, i)}$ are acquired from excitation vector~$\M{V}^{(\alpha, i)}$ as
\begin{equation}
    \M{v}^{(\alpha, i)} = \left( \M{P}^\herm{} \M{P} \right)^{-1} \M{P}^\herm{} \M{V}^{(\alpha, i)},
    \label{eq:v_AlphaFromV_Alpha}
\end{equation}
see~\eqref{eq:vFromV} in~Appendix~\ref{sec:ap-tarc}.

Being now equipped with symmetry-adapted excitation vectors~$\M{V}^{(\alpha, i)}=\M{P}\M{v}^{(\alpha, i)}$, constraints of~\eqref{eq:optimProblem} are automatically fulfilled irrespective of their number. The variables remaining for optimization~\eqref{eq:optimProblem} are therefore positions~$\xi$ of ports in the generator of the structure and their amplitudes. In a simplified case, when only one port exists in the generator of the structure, its amplitude is of no relevance and the only optimized variable is position~$\xi$, \ie{}, the optimization problem~\eqref{eq:optimProblem} reduces to 
\begin{equation}
        \underset{\xi}{\T{minimize}} \quad t_{\T{RMS}}.
    \label{eq:optimProblem2}
\end{equation}

In order to give a simple set of instructions for the procedure above, the TARC minimization with fully orthogonal states iteratively performs:
\begin{enumerate}
    \item Pick a position $\xi$.
    \item Create a port matrix $\M{P}$, see~\eqref{eq:VbyVp} in~Appendix~\ref{sec:ap-tarc}.
    \item Construct vector~$\M{v}$ exciting only the ports in the generator of the structure.
    \item Perform symmetry-adaptation~\eqref{eq:symmetryAdaptedVec} of the vector $\M{V} = \M{P}\M{v}$ into all species $\left(\alpha, i\right)$.
    \item Get $\M{v}^{\left(\alpha, i\right)}$ via~\eqref{eq:v_AlphaFromV_Alpha} for each species.
    \item Calculate \ac{TARC} $t\left(\M{v}^{\left(\alpha, i\right)}\right)$ for all species~\eqref{eq:TARC:Vp}.
    \item Evaluate the fitness function~$t_\T{RMS}$ via~\eqref{eq:RMS_freq}.
\end{enumerate}

\subsection{Single-Frequency Analysis}
\label{sec:example:singleFreq}
The optimization of the port's placement in the generator of the structure \eqref{eq:optimProblem2} computed at the single frequency sample $\left(N_\T{f} = 1\right)$ is analyzed in this section. The selected frequency corresponds to the antenna's electrical size $ka = 10.19$. The low number of tested positions and the possibility to precalculate all matrix operators $\M{A}$ enables the use of an extensive search to evaluate~\eqref{eq:TARC:Vp} for each tested position $\xi$ depicted by the red color in Fig.~\ref{fig:rectangularRim:mesh}.

\begin{figure}[t]
    \centering
    \includegraphics[width = 0.5\textwidth]{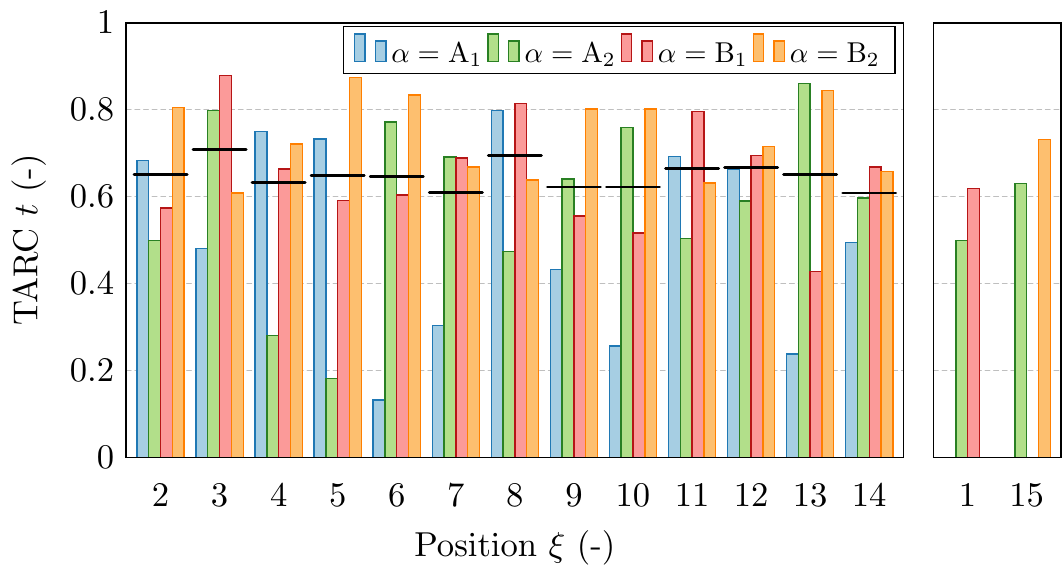}
    \caption{\ac{TARC} values of four orthogonal states $\left(\alpha, 1\right)$ for the rectangular rim from Fig.~\ref{fig:rectangularRim:mesh} evaluated for different positions of port in the generator of the structure $\xi$ at $ka = 10.19$. Black lines denote \ac{RMS} values $t_{\T{RMS}}$. The states are named according to irreducible representations, see Table~\ref{tab:placementCollision}.}
    \label{fig:rectangularRim:TARC_BF}
\end{figure}

The results are presented in Fig.~\ref{fig:rectangularRim:TARC_BF}. As mentioned in Section~\ref{sec:ExcitationStatesPG:badPlacing}, positions $\xi = 1$ and $\xi = 15$ are not able to excite all four orthogonal states since they are placed at the reflection plane. All other positions~$\xi$ result in a total of four symmetrically placed ports providing four orthogonal states.

Bars in Fig.~\ref{fig:rectangularRim:TARC_BF} show \ac{TARC} values~\eqref{eq:TARC:Vp} computed for each of the four species $(\alpha,1)$, $\alpha \in \left\lbrace \T{A}_1, \T{A}_2, \T{B}_1, \T{B}_2\right\rbrace$. The values~$t_{\T{RMS}}$ are represented by the black vertical lines. The optimal port position $\xi$ in the generator of the structure is declared as the one with the lowest value of~$t_{\T{RMS}}$, \ie{}, position~$\xi = 14$.

\begin{figure}[t]
    \centering
    \includegraphics[width = 8.9cm]{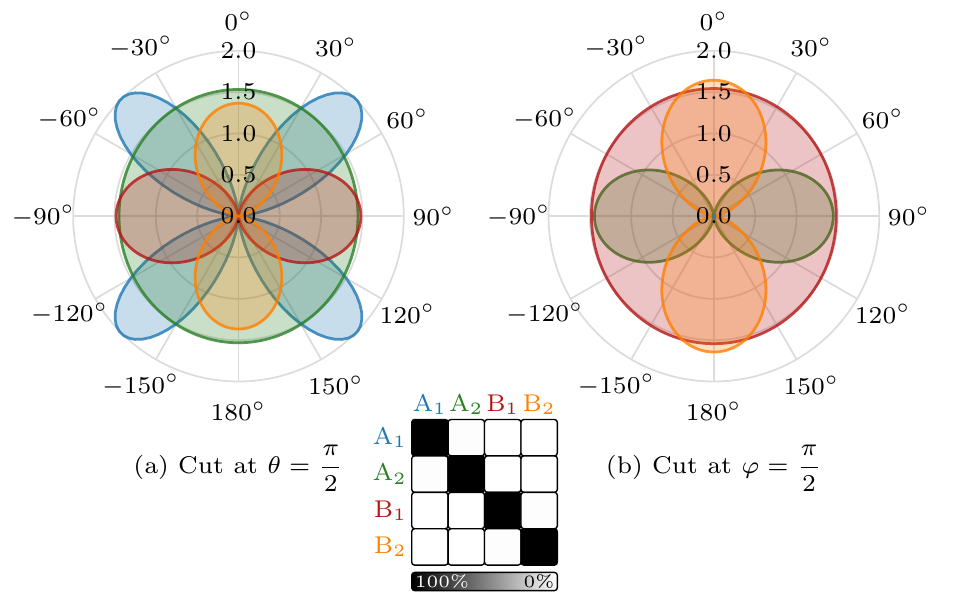}
    \caption{Far-field cuts with polarization along direction~$\UV{\varphi}$ computed at $ka = 1$ for  excitation vectors $\M{V}^{\left(\alpha, i\right)}$ for $\xi=7$. Radiation patterns are orthogonal which is confirmed by the envelope correlation coefficient~\cite{RosengrenKildal_RadiationEfficiencyCorrelationDiversityForMIMO} depicted in the table. The naming convention adapted for the states is the same as in Fig.~\ref{fig:rectangularRim:TARC_BF} and in Table~\ref{tab:placementCollision}.}
    \label{fig:rectangularRim_FF_cuts}
\end{figure}

Radiated patterns were computed and plotted as two-dimensional cuts in Fig.~\ref{fig:rectangularRim_FF_cuts} to confirm the orthogonality of the designed excitation vectors $\M{V}^{\left(\alpha, i\right)}\left(\xi\right)$. One can see that these patterns are similar to spherical harmonics which are orthogonal~\cite{Hamermesh_GroupTheoryandItsApplicationToPhysicalProblems_1989}. To reduce the complexity of radiation patterns in Fig.~\ref{fig:rectangularRim_FF_cuts}, radiation patterns were computed at $ka = 1$, bearing in mind that the orthogonality between states is frequency independent.

\subsection{Frequency Range Analysis}
\label{sec:example:freqRange}
\begin{figure}[t]
    \centering
    \includegraphics[width = 8.9cm]{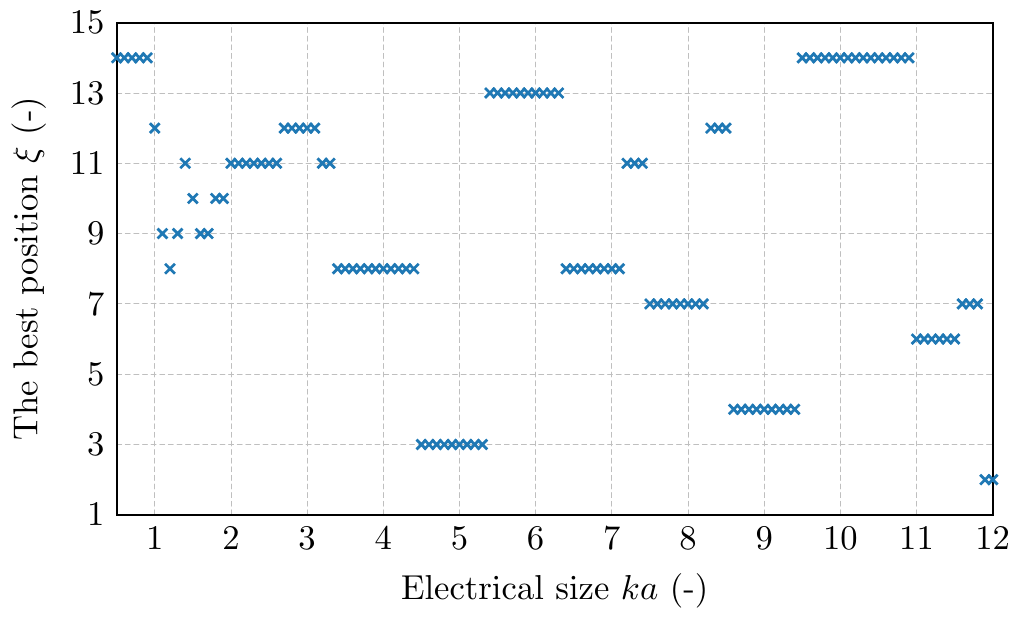}
    \caption{The best position of the port in the generator of the structure~$\xi$ with respect to \ac{TARC} value~$t_{\T{RMS}}$ evaluated at each frequency sample.}
    \label{fig:rectangularRim_bestPosOverFreq}
\end{figure}

Multiport antenna systems typically operate in a wide frequency range. However, evaluating~\eqref{eq:optimProblem2} at each frequency, as was done in the previous section, does not provide a unique best position~$\xi$, see Fig.~\ref{fig:rectangularRim_bestPosOverFreq}, where $N_\T{f}=116$ frequency samples in the range corresponding to the antenna's electrical size $ka \in \left(0.5,12\right)$ was used.

\begin{figure}[t]
    \centering
    \includegraphics[width = 8.9cm]{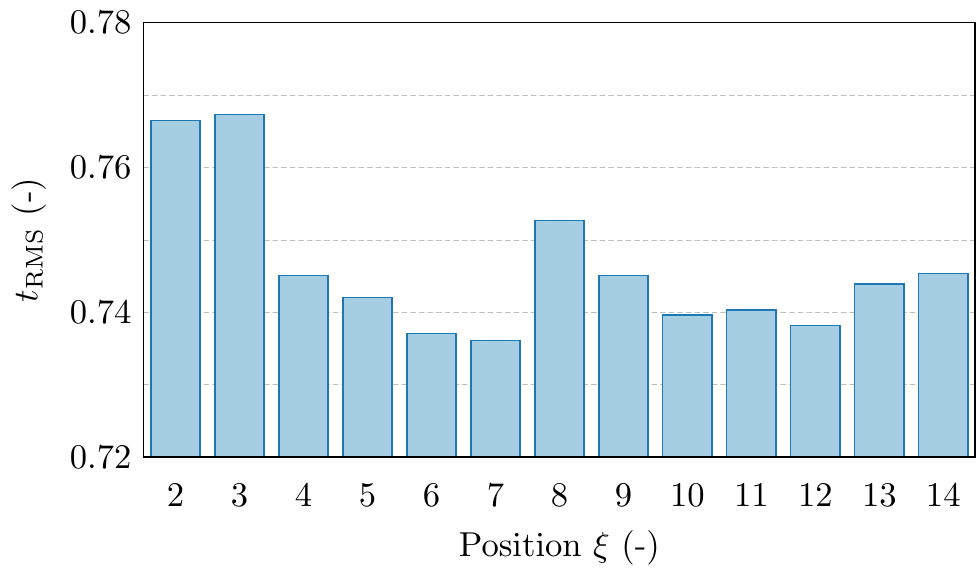}
    \caption{\ac{TARC} values evaluated by~\eqref{eq:RMS_freq} for different positions $\xi$ of the sole ($N_\xi = 1$) feeder which was symmetry-adapted.}
    \label{fig:rectangularRim_RMSfreqOverBF}
\end{figure}

\begin{figure}[t]
    \centering
    \includegraphics[width = 8.9cm]{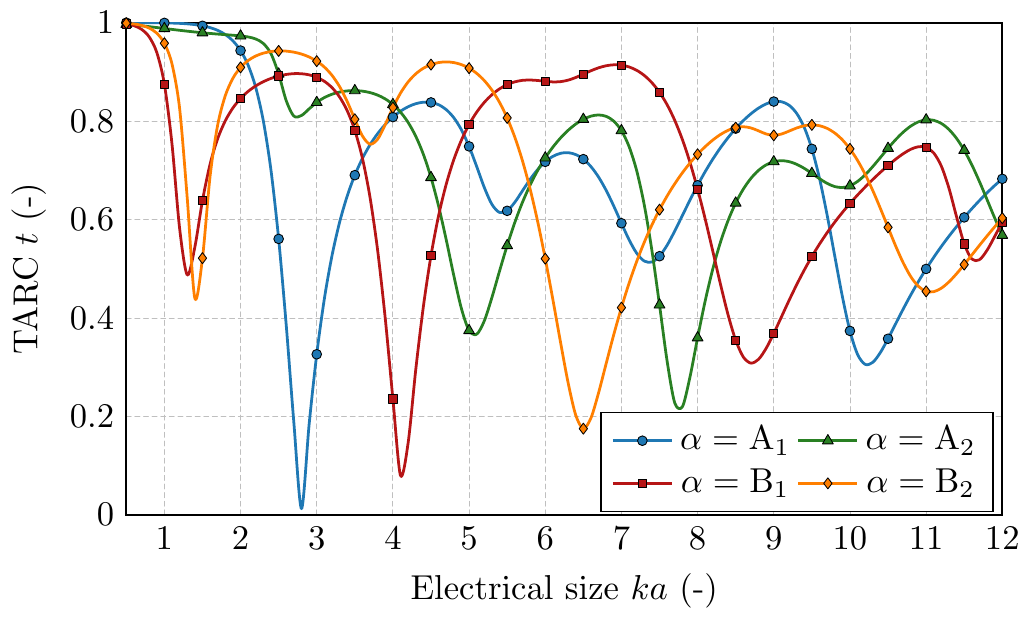}
    \caption{\ac{TARC} values of four orthogonal states $\left(\alpha, 1\right)$ for a rectangular rim depicted in Fig.~\ref{fig:rectangularRim:mesh}, the best position $\xi = 7$ is considered.}
    \label{fig:rectangularRim_TARCoverKa}
\end{figure}

The unique solution is accomplished by evaluating the \ac{RMS} value of \ac{TARC}~\eqref{eq:RMS_freq} over the frequency range in which the best position minimizing \eqref{eq:optimProblem2} is $\xi=7$, see Fig.~\ref{fig:rectangularRim_RMSfreqOverBF}. The realized \ac{TARC} computed for this optimal position over the whole frequency band is shown in Fig.~\ref{fig:rectangularRim_TARCoverKa}. It can be observed that there is no frequency where all four states radiate well, which results from their different current distributions. However, minimizing~\eqref{eq:optimProblem2}, by counting all frequencies of the selected band, provides a solution in which average \ac{TARC} over all channels is the best.

The values in Fig.~\ref{fig:rectangularRim_RMSfreqOverBF} are not so different and, in fact, are unsatisfactory. This is caused by the wide frequency range used and by employing the connected transmission lines of characteristic impedance~$Z_0 = 50\,\Omega$, which is not an optimal value for the chosen structure. Optimization of the impedance matching would demand a topological change of the antenna structure (keeping the necessary symmetries) which is beyond the scope of this paper.

\subsection{More Ports Placed in the Generator of the Structure}
\label{sec:MorePorts}
The previous subsections assumed the existence of a sole port placed in the generator of the structure which was symmetry-adapted. Nevertheless, a higher number of ports might give better radiation properties. In the case of~\mbox{$N_\xi > 1$} ports placed in the generator of the structure, in addition to all statements in Section~\ref{sec:optimProblem}, the complex amplitudes connected to the ports also have significance.

As port excitation vector $\M{v}$ is constructed so that only ports placed in the generator of structure are excited (\ie{}, there is $N_\xi$ nonzero values) and because the symmetry-adaptation process~\eqref{eq:symmetryAdaptedVec} transforms these $N_\xi$ nonzero values to $N_\T{p}$ nonzero values in vector $\M{v}^{\left(\alpha, i\right)}$, the symmetry-adapted vector can also be expressed as
\begin{equation}
    \M{v}^{\left(\alpha, i\right)} = \M{p}^{\left(\alpha, i\right)} \V{\kappa}^{\left(\alpha, i\right)},
    \label{eq:vAsKappa}
\end{equation}
where~$\M{p}\in \mathbb{R}^{N_\T{p}\times N_\xi}$ is a port-indexing matrix (each column in~$\M{p}$ corresponds to one exclusively excited port in the generator of the structure) and vector $\V{\kappa}$ of size $N_\xi\times 1$ contains only voltages of ports placed in the generator of the structure.

Substituting \eqref{eq:vAsKappa} into \eqref{eq:TARC:Vp} leads to
\begin{equation}
    \label{eq:TARC:kappa}
    t^{\left(\alpha, i\right)}\left(\V{\kappa}^{\left(\alpha, i\right)}\right) = \sqrt{1 - \frac
    {\left(\V{\kappa}^{\left(\alpha, i\right)}\right)^\herm{} \M{A}^{\left(\alpha, i\right)} \V{\kappa}^{\left(\alpha, i\right)}}
    {\left(\V{\kappa}^{\left(\alpha, i\right)}\right)^\herm{} \M{B}^{\left(\alpha, i\right)} \V{\kappa}^{\left(\alpha, i\right)}}},
\end{equation}
where
\begin{equation}
    \M{A}^{\left(\alpha, i\right)} = 4 Z_0 \left(\M{Y} \M{P} \M{p}^{\left(\alpha, i\right)}\right)^\herm{} \M{R}_0 \M{Y} \M{P} \M{p}^{\left(\alpha, i\right)} \quad \in\mathbb{C}^{N_\xi \times N_\xi}
\end{equation}
and
\begin{equation}
    \M{B}^{\left(\alpha, i\right)} = \left(\M{k} \M{p}^{\left(\alpha, i\right)}\right)^\herm{}\M{k} \M{p}^{\left(\alpha, i\right)}\quad \in\mathbb{C}^{N_\xi \times N_\xi}.
\end{equation}

In order to minimize~\eqref{eq:TARC:kappa}, a generalized eigenvalue problem
\begin{equation}
    \label{eq:TARC:eig}
    \M{A}^{\left(\alpha, i\right)} \V{\kappa}_\T{p}^{\left(\alpha, i\right)} = \lambda_\T{p}^{\left(\alpha, i\right)} \M{B}^{\left(\alpha, i\right)} \V{\kappa}_\T{p}^{\left(\alpha, i\right)}
\end{equation}
is solved and an eigenvector minimizing~\eqref{eq:TARC:kappa}, \ie{}, one corresponding to the highest eigenvalue~$\lambda_\T{p}^{\left(\alpha, i\right)}$, is chosen. This solution provides the best achievable \ac{TARC} for a given species $\left(\alpha, i\right)$, the value of which is
\begin{equation}
    t_{\T{bound}}^{\left(\alpha, i\right)} = \sqrt{1-\T{max}\left(\lambda_\T{p}^{\left(\alpha, i\right)}\right)}.
\end{equation}

In the case of more ports placed in the generator of the structure, the process described in this section must be used in every step of optimization~\eqref{eq:optimProblem2}, \ie{}, vectors~$\V{\kappa}^{\left(\alpha, i\right)}$ must be evaluated for each choice of~$\xi$.

\subsection{Analysis with More Ports in the Generator of the Structure}
\label{sec:MorePortsAnalysis}
The optimal placement of two and three ports, $N_\xi = \left\lbrace2,3\right\rbrace$, in the generator of the structure is studied in this section. The same metallic rim as in Section~\ref{sec:example:singleFreq} operating at $ka = 10.19$ is used and the method from Section~\ref{sec:MorePorts} is applied, see Table~\ref{tab:rectangularRim_TARCoptim} for the results. It can be observed that the involvement of more ports significantly decreases the \ac{RMS} of \ac{TARC} across the states. This is because the optimal current density reaching minimal \ac{TARC} is better approximated with more excitation ports.

Table~\ref{tab:rectangularRim_TARCoptim} shows results for a port configuration adopted from~\cite{PeitzmeierManteuffel_SelectiveExcitationCMs_EuCAP} which was discussed in Section~\ref{sec:illustrativeExample}. This configuration uses $N_\xi=3$ ports placed in the generator of the structure and $N_\T{p} = 8$ ports. Nevertheless, Table~\ref{tab:rectangularRim_TARCoptim} reveals that better results may be obtained when the symmetry-adapted basis described in this paper is utilized.

\begin{table}[t] 
    \centering 
    \caption{The best values of \ac{RMS} of \ac{TARC} for various excitation strategies on a rectangular rim depicted in Fig.~\ref{fig:rectangularRim:mesh}.}
    \begin{tabular}{ccccc}
        Solution & $N_\xi$ & $\left\lbrace\xi_\T{p}\right\rbrace$ &$N_\T{p}$ &  $t_{\T{RMS}}$\\[0.1cm] \toprule
        Best $1$& $1$ & $14$ & $4$ & $0.608$\\
        Best $2$& $2$ & $10,~11$ & $8$ & $0.400$\\
        Best $3$& $3$ & $11,~12,~13$ & $12$ & $0.317$\\
        Table~\ref{tab:dirkPlateVoltage}, column $2$ & $3$ & $1,~10,~15$ & $8$ & $0.587$\\
         \bottomrule
    \end{tabular} 
    \label{tab:rectangularRim_TARCoptim}
\end{table}

\begin{figure}[t]
    \centering
    \includegraphics[width = 8.9cm]{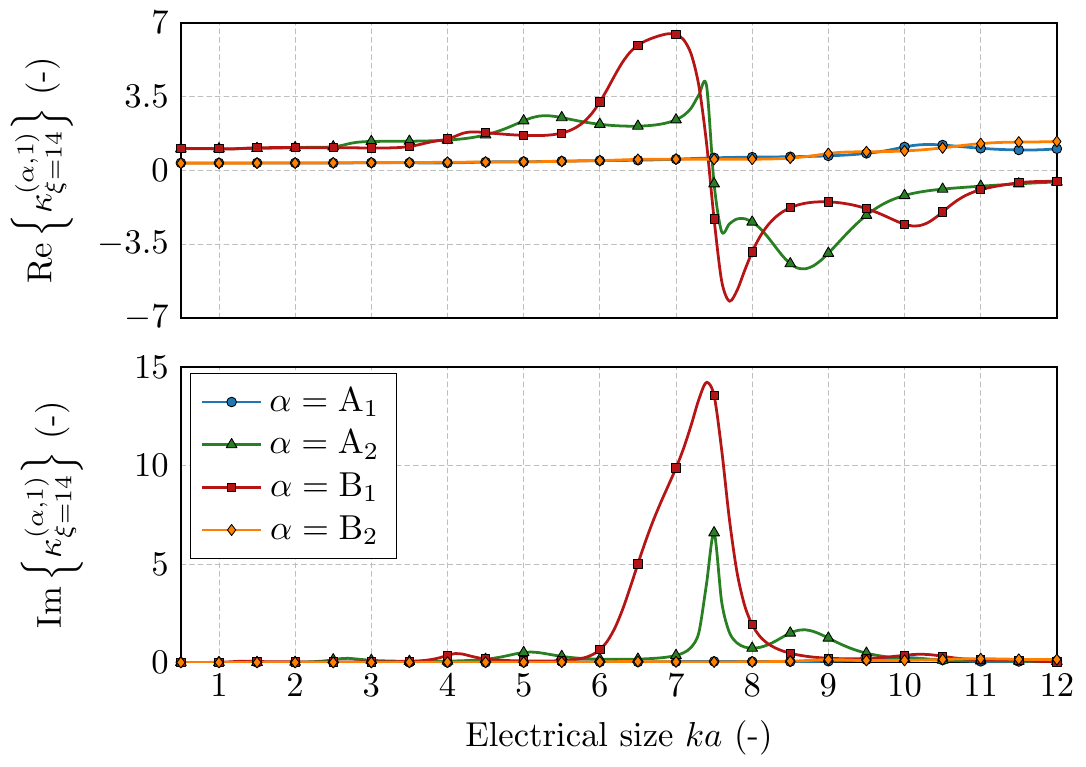}
    \caption{Voltage amplitudes $\V{\kappa}^{\left(\alpha, 1\right)}$ for a configuration with a combination of $N_\xi = 2$ positions at $\xi \in \left\lbrace 12, 14\right\rbrace$. The voltage impressed to the port at position~$\xi = 12$ is normalized to one volt.}
    \label{fig:rectangularRim_twoPorts-voltages}
\end{figure}

\begin{figure}[t]
    \centering
    \includegraphics[width = 8.9cm]{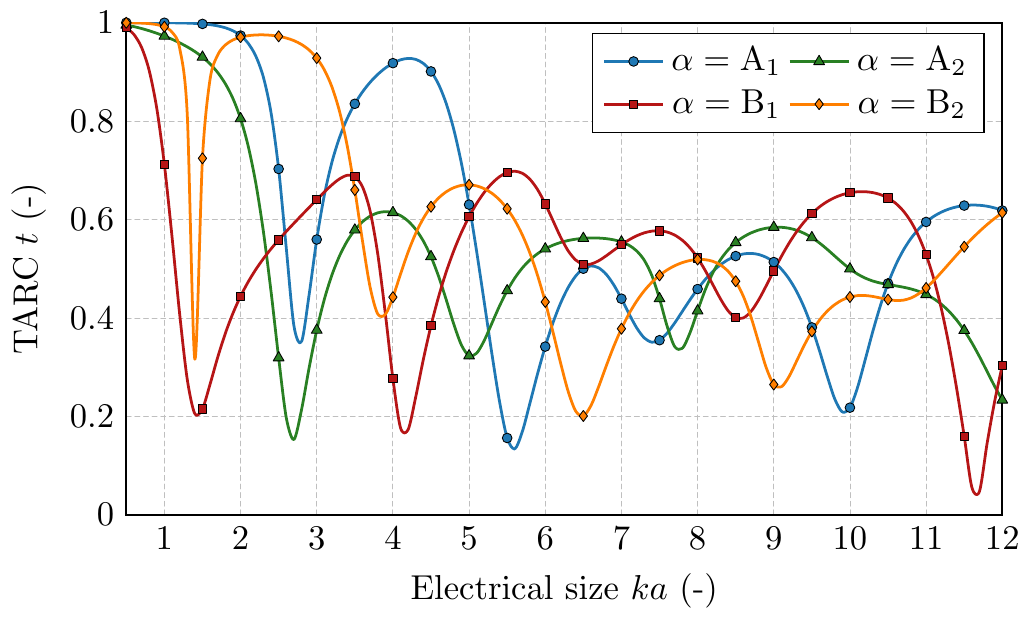}
    \caption{\ac{TARC} values of four orthogonal states $\left(\alpha, 1\right)$ for the rectangular rim depicted in Fig.~\ref{fig:rectangularRim:mesh} and a combination of $N_\xi = 2$ positions at $\xi \in \left\lbrace 12, 14\right\rbrace$.}
    \label{fig:rectangularRim_twoPorts-TARCoverKa}
\end{figure}

The frequency range analysis from Section~\ref{sec:example:freqRange} was repeated for the combination of $N_\xi = 2$ ports placed in the generator of the structure. Ports at positions $\xi \in \left\lbrace 12, 14\right\rbrace$ provide the lowest \ac{RMS} \eqref{eq:RMS_freq} $t_{\T{RMS}} = 0.605$. However, the solution with a combination of more positions $\left\lbrace\xi_k\right\rbrace$ requires optimized port voltage amplitudes $\V{\kappa}$~\eqref{eq:vAsKappa} which vary over frequency, see~Fig.~\ref{fig:rectangularRim_twoPorts-voltages}.
Figure~\ref{fig:rectangularRim_twoPorts-TARCoverKa} shows realized \ac{TARC} values reached by this configuration. The radiation efficiency is significantly improved as compared to the previous solution shown in Fig.~\ref{fig:rectangularRim_TARCoverKa}.

\section{Conclusion}
\label{sec:conclusion}
The presence of symmetries was utilized via point group theory to describe a procedure that determines where to place ports on an antenna to achieve orthogonal states with respect to any radiation metric, such as radiation and total efficiency, antenna gain, or Q-factor.

The methodology can play an essential role in the design of MIMO antennas when a few ports can orthogonalize several states (\eg{}, four ports on a rectangular structure generate four orthogonal states, eight ports on a square structure generate six orthogonal states, etc.). The maximal number of orthogonal states and the minimal number of ports needed to excite all of them is determined only from the knowledge of the point group to which the given geometry belongs. Due to the symmetries, the procedure of ports' placement can be accelerated by the reduction of the section where the port placed in the region of the generator of the structure can be placed and subsequently \Quot{symmetry-adapted} to the proper positions at the entire structure. It was also demonstrated that port positions intersecting reflection planes should not be used since they do not allow the excitation of all states.


A proper placement of ports was illustrated by an example -- with a single frequency and frequency range analysis -- featuring a simultaneous minimization of \acl{TARC} across the realized orthogonal states. Leaving aside the final matching optimization, it has been clearly presented how symmetries can be utilized in the design of a multi-port antenna.

\appendices

\section{Matrix Operators}
\label{sec:ap-operators}

Many antenna metrics are expressible as quadratic forms over time-harmonic current density~$\V{J}(\V{r})$, \cite{GustafssonTayliEhrenborgEtAl_AntennaCurrentOptimizationUsingMatlabAndCVX, JelinekCapek_OptimalCurrentsOnArbitrarilyShapedSurfaces}, which is represented in a suitable basis~$\{\basisFcn_n(\V{r})\}$ as
\begin{equation}
    \boldsymbol{J} \left( \boldsymbol{r} \right) \approx \sum\limits_{n=1}^{N_\T{u}} I_n \boldsymbol{\psi}_n \left(\boldsymbol{r} \right)
    \label{eq:Jbasis}
\end{equation}
with $N_\T{u}$ being the number of basis functions. The metric~$p$ is then given as
\begin{equation}
    p = \langle \V{J}, \OP{A} \V{J} \rangle \approx \Ivec^\herm \left[ \langle \basisFcn_m (\V{r}), \OP{A} \basisFcn_m (\V{r}) \rangle \right] \Ivec = \Ivec^\herm \M{A} \Ivec.
\end{equation}
For example, the complex power balance~\cite{Balanis_Wiley_2005} for radiator~$\srcRegion$ made of a good conductor reads
\begin{equation}
    \Prad + P_\T{L} + 2 \J \omega \left( W_\T{m} - W_\T{e} \right) \approx \dfrac{1}{2} \Ivec^\herm \left( \M{Z}_0 + \M{R}_\rho \right)\Ivec,
\end{equation}
where the vacuum impedance matrix~$\M{Z}_0 = \M{R}_0 + \J \M{X}_0$ is defined element-wise as
\begin{equation}
    Z_{0,mn} = - \J \omega \MUE \int\limits_{\varOmega} \int\limits_{\varOmega} \basisFcn_m \left(\V{r} \right) \cdot \V{G} \left(\V{r}, \V{r} ' \right) \cdot \basisFcn_n \left(\V{r} ' \right) \D{S} \D{S '},
\end{equation}
with $\omega$ being angular frequency, $\MUE$ being vacuum permeability, and $\V{G}$ being free-space dyadic Green's function~\cite{VolakisSertel_IntegralEquationMethodsForElectromagnetics}. Ohmic losses~$P_\T{L}$ are represented via matrix~$\M{R}_\rho$ which, under thin-sheet approximation~\cite{SenoirVolakis_ApproximativeBoundaryConditionsInEM}, is defined element-wise as~\cite{JelinekCapek_OptimalCurrentsOnArbitrarilyShapedSurfaces}
\begin{equation}
    R_{\rho, mn} = \int\limits_\srcRegion \rho \left(\V{r}\right) \basisFcn_m (\V{r}) \cdot \basisFcn_n (\V{r}) \D{S},
\end{equation}
where~$\rho$ is surface resistivity~\cite{SenoirVolakis_ApproximativeBoundaryConditionsInEM}.
Another notable operator~\cite{Vandenbosch_ReactiveEnergiesImpedanceAndQFactorOfRadiatingStructures, CismasuGustafsson_FBWbySimpleFreuqSimulation}
\begin{equation}
    \M{W} = \omega\dfrac{\partial\M{X}_0}{\partial\omega}
    \label{eq:app1Q4}
\end{equation}
gives energy stored in the near-field on a device, thus determining the bandwidth potential of a radiator~\cite{CapekJelinekHazdra_OnTheFunctionalRelationBetweenQfactorAndFBW}.

\section{Excitation Vector}
\label{sec:ap-excitationVector}
The excitation of obstacle~$\srcRegion$ is realized by an incident electric field intensity~$\V{E}^\T{i} \left( \V{r} \right)$ represented element-wise in a basis~\eqref{eq:Jbasis} as
\begin{equation}
V_n = \int\limits_\srcRegion \basisFcn_n \left( \V{r} \right) \cdot \V{E}^\T{i} \left( \V{r} \right) \D{S},
\label{eq:Vvec}
\end{equation}
with~$\Vvec = [V_n]$ called the excitation vector. Incident field~$\V{E}^\T{i} \left( \V{r} \right)$ can be non-zero everywhere (then the vector~$\Vvec$ generally contains non-zero entries everywhere, \eg{}, a plane wave), or in a limited region only (then vector~$\Vvec$ is sparse, \eg{}, a delta-gap generator or a coaxial probe).

Considering electric field integral equation~\cite{Harrington_FieldComputationByMoM} in algebraic representation~\eqref{eq:Jbasis}, current solution~$\Ivec$ to a problem of given excitation~$\Vvec$ reads
\begin{equation}
\Zmat \Ivec = \Vvec,
\label{eq:ZIV}
\end{equation}
where $\Zmat = \M{Z}_0 + \M{R}_\rho$ is the system (impedance) matrix.

\section{Symmetry-Adaptation of a Vector}
\label{sec:ap-symmetryAdaptedVector}
\begin{figure}
    \centering
    \def\tmpHeight{3.5cm}
    \subfloat[]{\includegraphics[height = \tmpHeight]{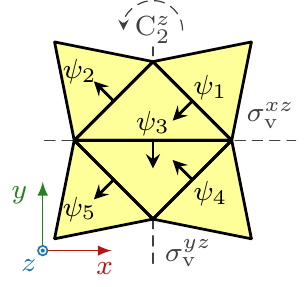}}~
    \subfloat[]{\includegraphics[height = \tmpHeight]{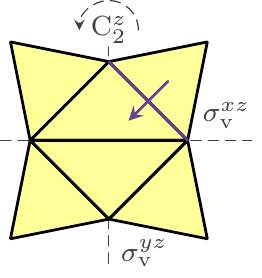}}\\
    \subfloat[]{\includegraphics[height = \tmpHeight]{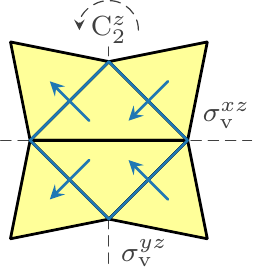}}~
    \subfloat[]{\includegraphics[height = \tmpHeight]{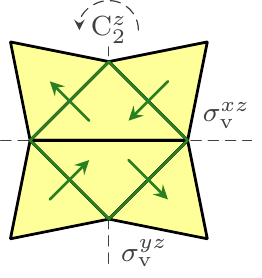}}\\
    \subfloat[]{\includegraphics[height = \tmpHeight]{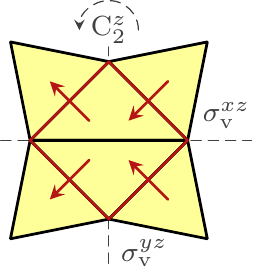}}~
    \subfloat[]{\includegraphics[height = \tmpHeight]{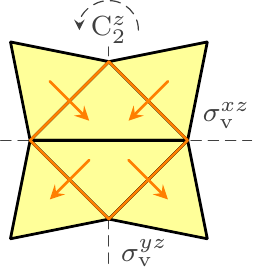}}
    \caption{(a) Five basis functions and their orientation on a star structure. (b) An excitation vector $\M{V}(1) = \left[1,0,0,0,0\right]^\trans$ was symmetry-adapted by \eqref{eq:symmetryAdaptedVec} to four irreps: (c)~$\alpha = \T{A}_1$, (d)~$\alpha = \T{A}_2$, (e)~$\alpha = \T{B}_1$, (f)~$\alpha = \T{B}_2$.}
    \label{fig:starStructure}
\end{figure}

The process of symmetry-adaptation of a vector \eqref{eq:symmetryAdaptedVec} is illustrated and explained in the example of a simple structure consisting of five RWG~\cite{RaoWiltonGlisson_ElectromagneticScatteringBySurfacesOfArbitraryShape} basis functions, see Fig.~\ref{fig:starStructure}(a). The delta-gap ports are connected directly to the basis functions, \ie{}, ports' positions $\xi$ are identical to the numbering of basis functions. This structure belongs to the same point group~$\T{C}_{2\T{v}}$ as the rectangular plate introduced in Section~\ref{sec:illustrativeExample} and is thus invariant to the same four symmetry operations: identity $\left(\T{E}\right)$, rotation by $\pi$ around $z$ axis $\left(\T{C}_2^z\right)$ and two reflections by $xz$ and $yz$ planes $\left(\sigma^{xz}_\T{v}, \sigma^{yz}_\T{v}\right)$. The point group $\T{C}_{2\T{v}}$ consists of four irreps $\alpha \in \left\lbrace\T{A}_1, \T{A}_2, \T{B}_1, \T{B}_2\right\rbrace$, with dimensionality $g^{(\alpha)} = 1$ for each irrep $\alpha$.

Mapping matrix $\M{C}\left(R\right)$, for each symmetry operation~$R$, is constructed so as to interlink pairs of basis functions which are mapped onto each other (respecting their orientation) via a given symmetry operation. Mapping matrices for the star structure from Fig.~\ref{fig:starStructure}(a) read
\begin{align}
    \M{C}\left(\T{E}\right) &= \T{diag}\left(\left[+1, +1, +1, +1, +1\right]\right),\\
    \M{C}\left(\T{C}_2^z\right) &= \left[\begin{array}{*{20}{c}}
       0 &  0 &  0 &  0 & -1 \\
       0 &  0 &  0 & -1 &  0 \\
       0 &  0 & -1 &  0 &  0 \\
       0 & -1 &  0 &  0 &  0 \\
      -1 &  0 &  0 &  0 &  0 \\
    \end{array}\right],\\
    \M{C}\left(\sigma^{xz}_\T{v}\right) &= \left[\begin{array}{*{20}{c}}
       0 &  0 &  0 & +1 &  0 \\
       0 &  0 &  0 &  0 & +1 \\
       0 &  0 & -1 &  0 &  0 \\
      +1 &  0 &  0 &  0 &  0 \\
       0 & +1 &  0 &  0 &  0 \\
    \end{array}\right],\\
    \M{C}\left(\sigma^{yz}_\T{v}\right) &= \left[\begin{array}{*{20}{c}}
       0 & -1 &  0 &  0 &  0 \\
      -1 &  0 &  0 &  0 &  0 \\
       0 &  0 & +1 &  0 &  0 \\
       0 &  0 &  0 &  0 & -1 \\
       0 &  0 &  0 & -1 &  0 \\
    \end{array}\right].
\end{align}

A general framework of how to obtain irreducible matrix representations $\M{D}^{(\alpha)}\left(R\right)$ is described in~\cite[Sec. II.B]{Maseketal_ModalTrackingBasedOnGroupTheory}. However, for one-dimensional irreps, the matrices~$\M{D}^{(\alpha)}\left(R\right)$ can be obtained directly from the character table, see the character table for the $C_{2\T{v}}$ point group in Table~\ref{tab:charTableC2v}. These character tables are known~\cite{McWeeny_GroupTheory} and unique for all point groups. For each irrep $\alpha$ (row) and each symmetry operation (column) the entry in the character table, called \Quot{a~character}, is $\chi^{(\alpha)}\left(R\right) = \T{trace}\left(\M{D}^{(\alpha)}\left(R\right)\right)$. Since the dimensionality of all irreps of the point group $C_{2\T{v}}$ is one ($g^{(\alpha)} = 1$ for each irrep $\alpha$), values in the character table are equal to the irreducible matrix representations~$\M{D}^{(\alpha)}\left(R\right)$ (matrices of size $1\times 1$).

\begin{table}[]
\centering
\caption{Character table for point group~$\T{C}_\T{2v}$ \cite{McWeeny_GroupTheory}, a rectangular plate and the star structure belong to.}
\begin{tabular}{ccccc}
$\T{C}_\T{2v}$ & $\T{E}$ & $\T{C}_\T{2}^z$ & $\T{\sigma}^{xz}_\T{v}$ & $\T{\sigma}^{yz}_\T{v}$ \\ \toprule
$\textsc{A}_1$ & $+1$ & $+1$ & $+1$ & $+1$ \\
$\textsc{A}_2$ & $+1$ & $+1$ & $-1$ & $-1$ \\
$\textsc{B}_1$ & $+1$ & $-1$ & $+1$ & $-1$ \\
$\textsc{B}_2$ & $+1$ & $-1$ & $-1$ & $+1$ \\ \bottomrule
\end{tabular}
\label{tab:charTableC2v}
\end{table}

The position of the initial port~$\xi$ can be freely chosen within the generator of the structure. Let us pick the position at $\xi = 1$ and construct an excitation vector~$\M{V}(1) = \left[1,\,0,\,0,\,0,\,0\right]^\trans$, see Fig.~\ref{fig:starStructure}(b).

Once matrices $\M{C}\left(R\right)$ and $\M{D}^{(\alpha)}\left(R\right)$ are known, a symmetry-adaptation of the excitation vector $\M{V}(1)$ into a given species $(\alpha, i)$ can be processed. The equation \eqref{eq:symmetryAdaptedVec} can be read as: An initial port recorded in~$\M{V}\left(\xi\right)$ is mapped onto its \Quot{doublet} under symmetry operation $R$ via mapping matrix~$\M{C}\left(R\right)$ while multiplying by a proper value from matrix~$\M{D}^{(\alpha)}\left(R\right)$ (in this case only values~$\pm1$) adds and provides a orthogonality property to the final symmetry-adapted vector $\M{V}^{\left(\alpha, i\right)}$:
\begin{align}
    \M{V}^{\left(\T{A}_1, 1\right)} &= \left[+1, -1, 0, +1, -1\right]^\trans,\\
    \M{V}^{\left(\T{A}_2, 1\right)} &= \left[+1, +1, 0, -1, -1\right]^\trans,\\
    \M{V}^{\left(\T{B}_1, 1\right)} &= \left[+1, +1, 0, +1, +1\right]^\trans,\\
    \M{V}^{\left(\T{B}_2, 1\right)} &= \left[+1, -1, 0, -1, +1\right]^\trans.
\end{align}
These solutions are shown in Fig.~\ref{fig:starStructure}(c--f). The normalization $g^{(\alpha)}/g = 1/4$ is intentionally omitted for each of solutions.

\section{Total Active Reflection Coefficient}
\label{sec:ap-tarc}
In order to derive~\eqref{eq:TARC:Vp}, incident power~$P_\T{in}$ is written using incident power waves~$\M{a} \in \mathbb{C}^{P\times 1}$ at antenna ports~\cite{Pozar_MicrowaveEngineering} as 
\begin{equation}
    \label{eq:Pin:port}
    P_{\T{in}} = \frac{1}{2} \M{a}^\herm{} \M{a}
\end{equation}
and the radiated power is written as~\cite{Harrington_FieldComputationByMoM}
\begin{equation}
    \label{eq:Prad:MoM}
    P_{\T{rad}} = \frac{1}{2} \M{I}^\herm{} \M{R}_0 \M{I},
\end{equation}
where $\M{R}_0 \in \mathbb{R}^{N_\T{u}\times N_\T{u}}$ is a radiation part of impedance matrix $\M{Z} \in \mathbb{C}^{N_\T{u} \times N_\T{u}}$ and $\M{I} \in \mathbb{C}^{N_\T{u}\times 1}$ is a vector of expansion coefficients within the \ac{MoM} solution to the \ac{EFIE}~\cite{Harrington_FieldComputationByMoM}, see Appendix~\ref{sec:ap-operators}. Using \eqref{eq:ZIV} it holds that 
\begin{equation}
\label{eq:App:Prad}
    P_{\T{rad}} = \frac{1}{2} \M{V}^\herm{} \M{Y}^\herm{} \M{R}_0 \M{Y}\M{V}.
\end{equation}

Assume an antenna fed by ports connected to transmission lines of real characteristic impedance~$Z_0$. Within the \ac{MoM} paradigm~\cite{Harrington_FieldComputationByMoM}, the excitation vector is
\begin{equation}
    \label{eq:VbyVp}
    \M{V} = \M{P} \M{v},
\end{equation}
where $\M{v}$ are port voltages and matrix $\M{P} \in \mathbb{R}^{N_\T{u}\times N_\T{p}}$ is a matrix the columns of which are the representations of separate ports. Notice that
\begin{equation}
    \M{v} = \left( \M{P}^\herm{} \M{P} \right)^{-1} \M{P}^\herm{} \M{V}.
    \label{eq:vFromV}
\end{equation}
The incident power waves can be expressed as~\cite{Pozar_MicrowaveEngineering}
\begin{equation}
    \label{eq:defA}
    \M{a} = \frac{1}{2\sqrt{Z_0}} \left(\M{e} + Z_0 \M{y}\right) \M{v},
\end{equation}
where $\M{e}$ is an identity matrix and $\M{y}$ is the admittance matrix~\cite{Pozar_MicrowaveEngineering} for port-like quantities.

Substituting~\eqref{eq:App:Prad},~\eqref{eq:VbyVp} and~\eqref{eq:defA} into~\eqref{eq:TARC:def} results in~\eqref{eq:TARC:Vp}. More details about \ac{TARC} for multi-port lossy antennas can be found in~\cite{Capeketal_OptimalityOfTARCAndRealizedGainForMultiPortAntennas_Arxiv}.


\begin{IEEEbiography}[{\includegraphics[width=1in,height=1.25in,clip,keepaspectratio]{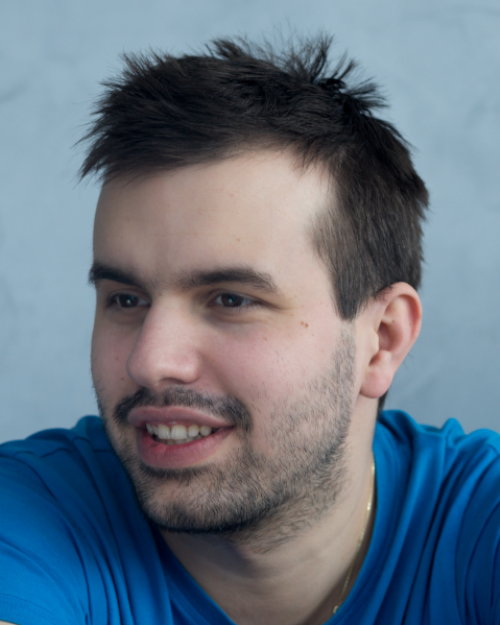}}]{Michal Masek}
received the M.Sc. degree in Electrical Engineering from Czech Technical University in Prague, Czech Republic, in 2015, where he is currently pursuing the Ph.D. degree in the area of modal tracking and characteristic modes. He is a member of the team developing the AToM (Antenna Toolbox for Matlab).
\end{IEEEbiography}

\begin{IEEEbiography}[{\includegraphics[width=1in,height=1.25in,clip,keepaspectratio]{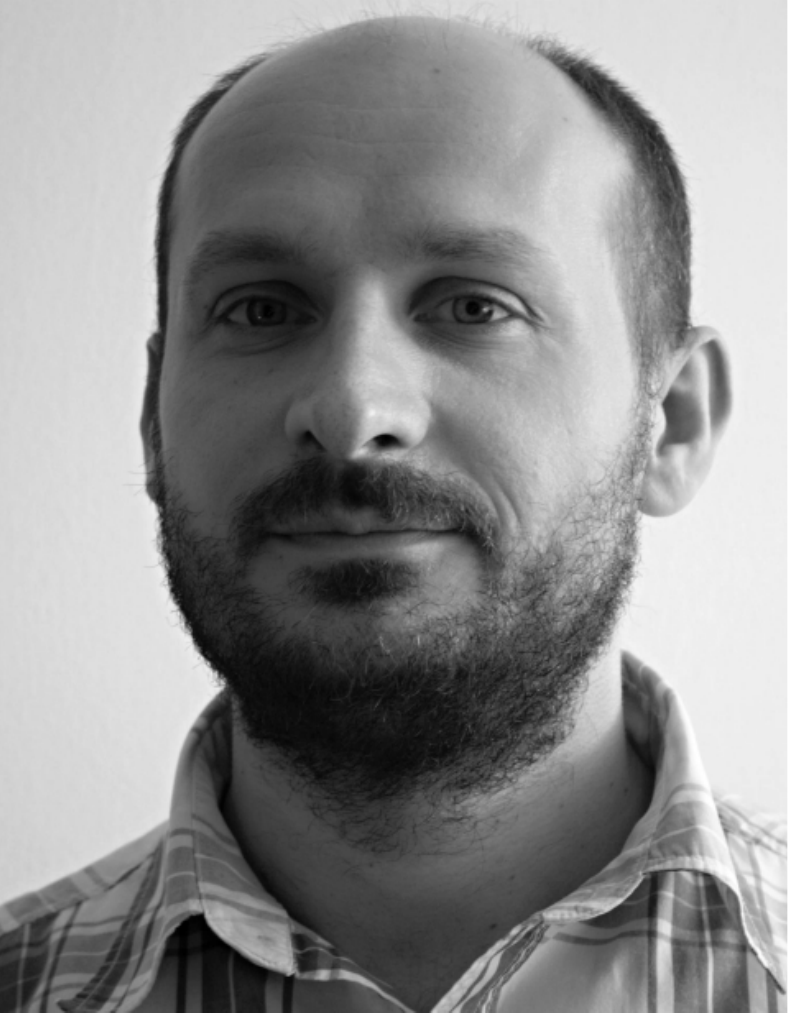}}]{Lukas Jelinek}
received his Ph.D. degree from the Czech Technical University in Prague, Czech Republic, in 2006. In 2015 he was appointed Associate Professor at the Department of Electromagnetic Field at the same university.

His research interests include wave propagation in complex media, general field theory, numerical techniques and optimization.
\end{IEEEbiography}

\begin{IEEEbiography}[{\includegraphics[width=1in,height=1.25in,clip,keepaspectratio]{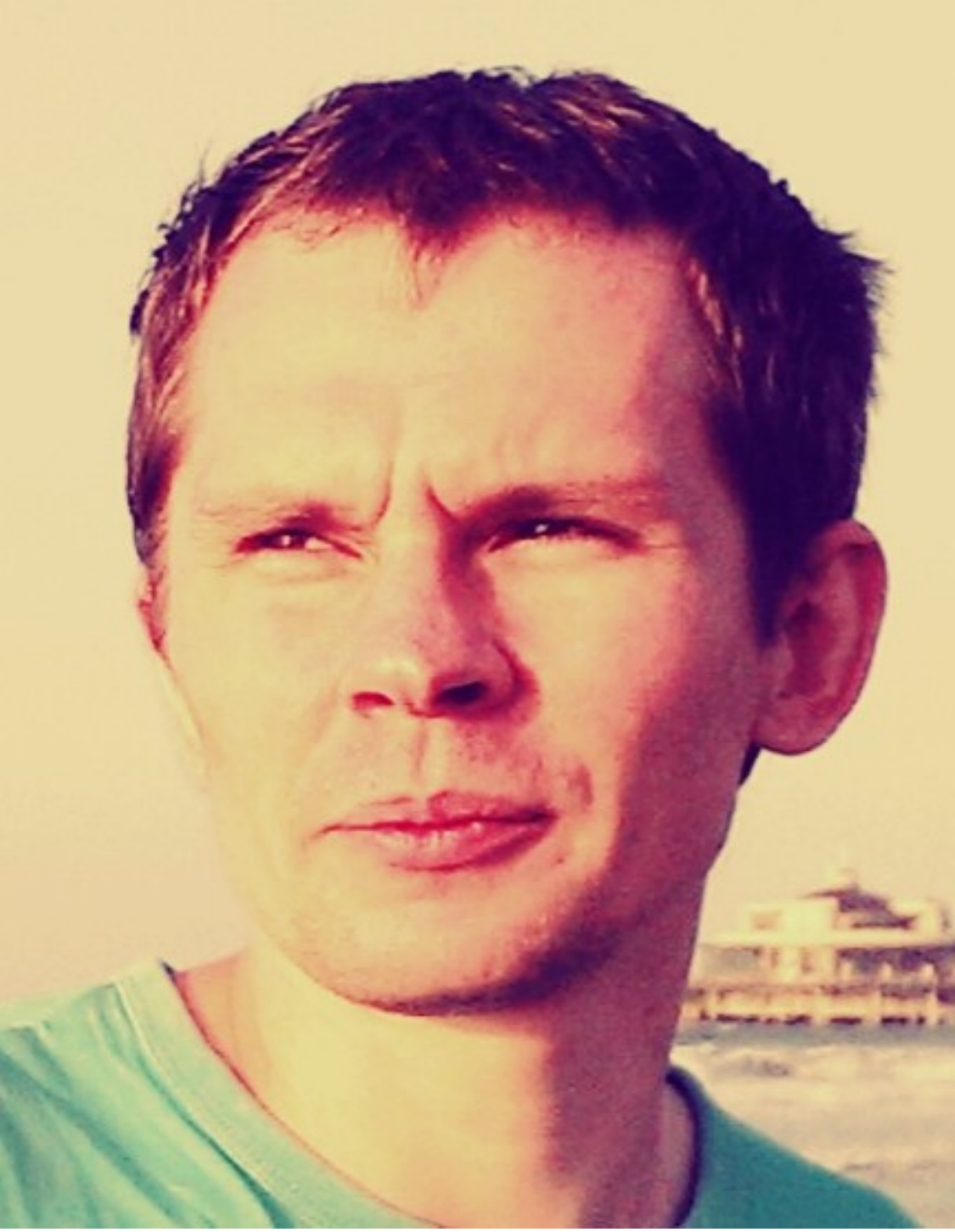}}]{Miloslav Capek}
(M'14, SM'17) received his M.Sc. degree in Electrical Engineering and Ph.D. degree from the Czech Technical University, Czech Republic, in 2009 and 2014, respectively. In 2017 he was appointed Associate Professor at the Department of Electromagnetic Field at the CTU in Prague.
	
He leads the development of the AToM (Antenna Toolbox for Matlab) package. His research interests are in the area of electromagnetic theory, electrically small antennas, numerical techniques, fractal geometry and optimization. He authored or co-authored over 100 journal and conference papers.

Dr. Capek is member of Radioengineering Society, regional delegate of EurAAP, and Associate Editor of IET Microwaves, Antennas \& Propagation.
\end{IEEEbiography}

\vspace{7pt}
\begin{thebibliography}{10}
	\providecommand{\url}[1]{#1}
	\csname url@samestyle\endcsname
	\providecommand{\newblock}{\relax}
	\providecommand{\bibinfo}[2]{#2}
	\providecommand{\BIBentrySTDinterwordspacing}{\spaceskip=0pt\relax}
	\providecommand{\BIBentryALTinterwordstretchfactor}{4}
	\providecommand{\BIBentryALTinterwordspacing}{\spaceskip=\fontdimen2\font plus
		\BIBentryALTinterwordstretchfactor\fontdimen3\font minus
		\fontdimen4\font\relax}
	\providecommand{\BIBforeignlanguage}[2]{{%
			\expandafter\ifx\csname l@#1\endcsname\relax
			\typeout{** WARNING: IEEEtran.bst: No hyphenation pattern has been}%
			\typeout{** loaded for the language `#1'. Using the pattern for}%
			\typeout{** the default language instead.}%
			\else
			\language=\csname l@#1\endcsname
			\fi
			#2}}
	\providecommand{\BIBdecl}{\relax}
	\BIBdecl
	
	\bibitem{GiordaniEtAl_Towards6GNetworks}
	\BIBentryALTinterwordspacing
	M.~Giordani, M.~Polese, M.~Mezzavilla, S.~Rangan, and M.~Zorzi, ``Towards {6G}
	networks: Use cases and technologies,'' 2020, eprint arXiv: 1903.12216.
	[Online]. Available: \url{https://arxiv.org/abs/1903.12216}
	\BIBentrySTDinterwordspacing
	
	\bibitem{Reed_NewTechnologiesSolvingSpectrumShortage_2016}
	J.~Reed, M.~Vassiliou, and S.~Shah, ``The role of new technologies in solving
	the spectrum shortage [point of view],'' \emph{Proceedings of the {IEEE}},
	vol. 104, no.~6, pp. 1163--1168, June 2016.
	
	\bibitem{GesbertEtAl_overviewMIMO}
	D.~Gesbert, M.~Shafi, D.~shan Shiu, P.~Smith, and A.~Naguib, ``From theory to
	practice: an overview of {MIMO} space-time coded wireless systems,''
	\emph{{IEEE} Journal on Selected Areas in Communications}, vol.~21, no.~3,
	pp. 281--302, Apr. 2003.
	
	\bibitem{Winter_CapacityOfRadioCommunationSystems}
	J.~Winters, ``On the capacity of radio communication systems with diversity in
	a rayleigh fading environment,'' \emph{{IEEE} Journal on Selected Areas in
		Communications}, vol.~5, no.~5, pp. 871--878, June 1987.
	
	\bibitem{JensenWallace_ReviewForMIMOWirelessCommunications}
	M.~Jensen and J.~Wallace, ``A review of antennas and propagation for {MIMO}
	wireless communications,'' \emph{{IEEE} Trans. Antennas Propag.}, vol.~52,
	no.~11, pp. 2810--2824, Nov. 2004.
	
	\bibitem{YangHanzo_FiftyYearsOfMIMO}
	S.~Yang and L.~Hanzo, ``Fifty years of {MIMO} detection: The road to
	large-scale {MIMOs},'' \emph{{IEEE} Communications Surveys {\&} Tutorials},
	vol.~17, no.~4, pp. 1941--1988, 2015.
	
	\bibitem{2010_Molisch_Book}
	A.~F. Molisch, \emph{Wireless Communications}.\hskip 1em plus 0.5em minus
	0.4em\relax Wiley-IEEE Press, 2010.
	
	\bibitem{AndersenRasmussen_DecouplingAndSescatteringNetworksForAntennas}
	J.~Andersen and H.~Rasmussen, ``Decoupling and descattering networks for
	antennas,'' \emph{{IEEE} Trans. Antennas Propag.}, vol.~24, no.~6, pp.
	841--846, Nov. 1976.
	
	\bibitem{Kildal_CorrelationAndCapacityOfMIMOSystems}
	P.-S. Kildal and K.~Rosengren, ``Correlation and capacity of {MIMO} systems and
	mutual coupling, radiation efficiency, and diversity gain of their antennas:
	simulations and measurements in a reverberation chamber,'' \emph{{IEEE}
		Communications Magazine}, vol.~42, no.~12, pp. 104--112, 2004.
	
	\bibitem{DaviuFabresGalloBataller_DesignOfAMultimodeMIMOantennaUsingCM}
	E.~Antonino-Daviu, M.~Cabedo-Fabres, M.~Gallo, M.~F. Bataller, and M.~Bozzetti,
	``Design of a multimode {MIMO} antenna using characteristic modes,'' in
	\emph{Proceedings of the 3rd European Conference on Antennas and Propagation
		(EUCAP)}, Berlin, Germany, Mar. 2009, pp. 1840--1844.
	
	\bibitem{Saarinen_etal_CombinatoryFeedingMethodForMobileApplication}
	T.~O. Saarinen, J.-M. Hannula, A.~Lehtovuori, and V.~Viikari, ``Combinatory
	feeding method for mobile applications,'' \emph{IEEE Antennas Wireless
		Propag. Lett.}, vol.~18, no.~7, pp. 1312--1316, July 2019.
	
	\bibitem{Hanulla_etal_FrequencyReconfiguragleMultibandHandset}
	J.-M. Hannula, T.~O. Saarinen, J.~Holopainen, and V.~Viikari, ``Frequency
	reconfigurable multiband handset antenna based on a multichannel
	transceiver,'' \emph{{IEEE} Trans. Antennas Propag.}, vol.~65, no.~9, pp.
	4452--4460, Sept. 2017.
	
	\bibitem{CoetzeeYu_PortDecouplingForSmallArraysByMeansOfAnEigenmodeFeedNetwork}
	J.~C. Coetzee and Y.~Yu, ``Port decoupling for small arrays by means of an
	eigenmode feed network,'' \emph{{IEEE} Trans. Antennas Propag.}, vol.~56,
	no.~6, pp. 1587--1593, June 2008.
	
	\bibitem{HarringtonMautz_TheoryOfCharacteristicModesForConductingBodies}
	R.~F. Harrington and J.~R. Mautz, ``Theory of characteristic modes for
	conducting bodies,'' \emph{IEEE Trans. Antennas Propag.}, vol.~19, no.~5, pp.
	622--628, Sept. 1971.
	
	\bibitem{HarringtonMautz_PatternSynthesisForLoadedNportScatterers}
	------, ``Pattern synthesis for loaded {N}-port scatterers,'' \emph{IEEE Trans.
		Antennas Propag.}, vol.~22, no.~2, pp. 184--190, 1974.
	
	\bibitem{PeitzmeierManteuffel_SelectiveExcitationCMs_EuCAP}
	N.~Peitzmeier and D.~Manteuffel, ``Selective excitation of characteristic modes
	on an electrically large antenna for mimo applications,'' in
	\emph{Proceedings of the 12th European Conference on Antennas and Propagation
		(EUCAP)}, 2018.
	
	\bibitem{Ethier_2009_TCM_MIMO_decoupled}
	J.~L.~T. Ethier and D.~McNamara, ``An interpretation of mode-decoupled {MIMO}
	antennas in terms of characteristic port modes,'' \emph{IEEE Trans. Magn.},
	vol.~45, no.~3, pp. 1128--1131, 2009.
	
	\bibitem{ChaudhurySchroederChaloupka_MIMOAntennaBasedOnOrthogonalityCMs}
	S.~K. Chaudhury, W.~L. Schroeder, and H.~J. Chaloupka, ``{MIMO} antenna system
	based on orthogonality of the characteristic modes of a mobile device,'' in
	\emph{2007 2nd International {ITG} Conference on Antennas}.\hskip 1em plus
	0.5em minus 0.4em\relax {IEEE}, Mar. 2007.
	
	\bibitem{LiShi_PatternRecinfigurableMIMOUsingCMs}
	K.~Li and Y.~Shi, ``A pattern reconfigurable {MIMO} antenna design using
	characteristic modes,'' \emph{{IEEE} Access}, vol.~6, pp. 43\,526--43\,534,
	2018.
	
	\bibitem{SuEtAl_RadiatonEnergyAndMutualCouplingForMIMOAntennaCMs}
	W.~Su, Q.~Zhang, S.~Alkaraki, Y.~Zhang, X.-Y. Zhang, and Y.~Gao, ``Radiation
	energy and mutual coupling evaluation for multimode {MIMO} antenna based on
	the theory of characteristic mode,'' \emph{{IEEE} Trans. Antennas Propag.},
	vol.~67, no.~1, pp. 74--84, Jan. 2019.
	
	\bibitem{LiMiersLau_DesignOfMIMOhandsetAntennasBasedOnCMmanipulationAtFrequencyBandsBelow1GHz}
	H.~Li, Z.~Miers, and B.~K. Lau, ``Design of orthogonal {MIMO} handset antennas
	based on characteristic mode manipulation at frequency bands below 1 {GHz},''
	\emph{IEEE Trans. Antennas Propag.}, vol.~62, no.~5, pp. 2756--5766, May
	2014.
	
	\bibitem{Manteuffel_Martens-CompactMultimodeMultielementAntennaForIndoorUWB}
	D.~Manteuffel and R.~Martens, ``Compact multimode multielement antenna for
	indoor {UWB} massive {MIMO},'' \emph{IEEE Trans. Antennas Propag.}, vol.~64,
	no.~7, pp. 2689--2697, July 2016.
	
	\bibitem{JaafarCollardeySharaiha_OptimizedmanipulationOfNetworkCMsForWidebandSmallAntennaMatching}
	H.~Jaafar, S.~Collardey, and A.~Sharaiha, ``Optimized manipulation of the
	network characteristic modes for wideband small antenna matching,''
	\emph{{IEEE} Transactions on Antennas and Propagation}, vol.~65, no.~11, pp.
	5757--5767, nov 2017.
	
	\bibitem{Wenetal_DesignOfMIMOAntennaForSmartwatchUsingTCM}
	D.~Wen, Y.~Hao, H.~Wang, and H.~Zhou, ``Design of a {MIMO} antenna with high
	isolation for smartwatch applications using the theory of characteristic
	modes,'' \emph{{IEEE} Transactions on Antennas and Propagation}, vol.~67,
	no.~3, pp. 1437--1447, Mar. 2019.
	
	\bibitem{QuetAl_MIMOAntennasUsingControlledOrthogonalCMs}
	L.~Qu, H.~Lee, H.~Shin, M.-G. Kim, and H.~Kim, ``{MIMO} antennas using
	controlled orthogonal characteristic modes by metal rims,'' \emph{{IET}
		Microwaves, Antennas {\&} Propagation}, vol.~11, no.~7, pp. 1009--1015, June
	2017.
	
	\bibitem{WuSuLiSu_ReductionInOutOfBandAntennaCouplingUsingCMs}
	Q.~{Wu}, W.~{Su}, Z.~{Li}, and D.~{Su}, ``Reduction in out-of-band antenna
	coupling using characteristic mode analysis,'' \emph{IEEE Transactions on
		Antennas and Propagation}, vol.~64, no.~7, pp. 2732--2742, July 2016.
	
	\bibitem{Ethier_2009_TCM_MIMO}
	J.~L.~T. Ethier and D.~McNamara, ``The use of generalized characteristic modes
	in the design of {MIMO} antennas,'' \emph{IEEE Trans. Magn.}, vol.~45, no.~3,
	pp. 1124--1127, 2009.
	
	\bibitem{GustafssonNordebo_CharacterizationOfMIMOAntennasUsingSVW}
	M.~Gustafsson and S.~Nordebo, ``Characterization of {MIMO} antennas using
	spherical vector waves,'' \emph{{IEEE} Transactions on Antennas and
		Propagation}, vol.~54, no.~9, pp. 2679--2682, sep 2006.
	
	\bibitem{GustafssonNordebo_SpectralEfficiencyOfSphere}
	------, ``On the spectral efficiency of a sphere,'' \emph{Progress In
		Electromagnetics Research}, vol.~67, pp. 275--296, 2007.
	
	\bibitem{XimenesAlmeida_CapacityOfMIMOSystemsUsingSWE}
	L.~R. Ximenes and A.~L.~F. de~Almeida, ``Capacity evaluation of {MIMO} antenna
	systems using spherical harmonics expansion,'' in \emph{2010 {IEEE} 72nd
		Vehicular Technology Conference - Fall}.\hskip 1em plus 0.5em minus
	0.4em\relax {IEEE}, sep 2010.
	
	\bibitem{GlazunovEtAl_PhysicalLimitationsOfInteractionOfSphereAndRandomField}
	A.~A. Glazunov, M.~Gustafsson, and A.~F. Molisch, ``On the physical limitations
	of the interaction of a spherical aperture and a random field,'' \emph{{IEEE}
		Transactions on Antennas and Propagation}, vol.~59, no.~1, pp. 119--128, jan
	2011.
	
	\bibitem{EhrenborgGustafsson_FundamentalBoundsOnMIMOAntennas}
	C.~Ehrenborg and M.~Gustafsson, ``Fundamental bounds on {MIMO} antennas,''
	\emph{{IEEE} Antennas and Wireless Propagation Letters}, vol.~17, no.~1, pp.
	21--24, jan 2018.
	
	\bibitem{Migliore_RoleOfNumberOfDOFinMIMOChannels}
	M.~Migliore, ``On the role of the number of degrees of freedom of the field in
	{MIMO} channels,'' \emph{{IEEE} Transactions on Antennas and Propagation},
	vol.~54, no.~2, pp. 620--628, feb 2006.
	
	\bibitem{GlazunovZhang2011_OptimalMIMOantennaCoefs}
	A.~A. Glazunov and J.~Zhang, ``On some optimal {MIMO} antenna coefficients in
	multipath channels,'' \emph{Progress In Electromagnetics Research B},
	vol.~35, pp. 87--109, 2011.
	
	\bibitem{Migliore_HorseIrMoreImporatntThanHorseman}
	M.~D. Migliore, ``Horse (electromagnetics) is more important than horseman
	(information) for wireless transmission,'' \emph{IEEE Trans. Antennas
		Propag.}, vol.~67, no.~4, pp. 2046--2055, April 2019.
	
	\bibitem{Mohajer2010_SWEforMIMOsystems}
	M.~Mohajer, S.~Safavi-Naeini, and S.~K. Chaudhuri, ``Spherical vector wave
	method for analysis and design of {MIMO} antenna systems,'' \emph{{IEEE}
		Antennas and Wireless Propagation Letters}, vol.~9, pp. 1267--1270, 2010.
	
	\bibitem{Capeketal_OptimalityOfTARCAndRealizedGainForMultiPortAntennas_Arxiv}
	\BIBentryALTinterwordspacing
	M.~Capek, L.~Jelinek, and M.~Masek, ``Finding optimal total active reflection
	coefficient and realized gain for multi-port lossy antennas,'' 2020, accepted
	to IEEE Trans. AP. [Online]. Available:
	\url{https://arxiv.org/abs/2002.12747}
	\BIBentrySTDinterwordspacing
	
	\bibitem{Mautz1973}
	J.~Mautz and R.~Harrington, ``Modal analysis of loaded {N}-port scatterers,''
	\emph{IEEE Trans. Antennas Propag.}, vol.~21, no.~2, pp. 188--199, Mar. 1973.
	
	\bibitem{WallaceJensen_MutualCouplingInMIMOWirelessSystems}
	J.~Wallace and M.~Jensen, ``Mutual coupling in {MIMO} wireless systems: A
	rigorous network theory analysis,'' \emph{{IEEE} Transactions on Wireless
		Communications}, vol.~3, no.~4, pp. 1317--1325, jul 2004.
	
	\bibitem{KrewskiSchroederSolbach_Multiband2portMIMOLTEantennaDesign}
	A.~Krewski, W.~L. Schroeder, and K.~Solbach, ``Multi-band 2-port {MIMO LTE}
	antenna design for laptops using characteristic modes,'' in \emph{Antennas
		and Propagation Conference (LAPC), Loughborough}, 2012, pp. 1--4.
	
	\bibitem{McWeeny_GroupTheory}
	R.~McWeeny, \emph{Symmetry: An Introduction to Group Theory and Its
		Applications}.\hskip 1em plus 0.5em minus 0.4em\relax London: Pergamon Press,
	1963.
	
	\bibitem{Hamermesh_GroupTheoryandItsApplicationToPhysicalProblems_1989}
	M.~Hamermesh, \emph{Group Theory and Its Application to Physical
		Problems}.\hskip 1em plus 0.5em minus 0.4em\relax Dover, 1989.
	
	\bibitem{SchabEtAl_EigenvalueCrossingAvoidanceInCM}
	K.~R. Schab, J.~M. Outwater~Jr., M.~W. Young, and J.~T. Bernhard, ``Eigenvalue
	crossing avoidance in characteristic modes,'' \emph{IEEE Trans. Antennas
		Propag.}, vol.~64, no.~7, pp. 2617--2627, July 2016.
	
	\bibitem{Knorr_1973_TCM_symmetry}
	J.~B. Knorr, ``Consequences of symmetry in the computation of characteristic
	modes for conducting bodies,'' \emph{IEEE Trans. Antennas Propag.}, vol.~21,
	no.~6, pp. 899--902, Nov. 1973.
	
	\bibitem{MartensManteuffel_SystematicDesignMethodOfMobileMultipleAntennaSystemUsingCM}
	R.~Martens and D.~Manteuffel, ``Systematic design method of a mobile multiple
	antenna system using the theory of characteristic modes,'' \emph{IET Microw.
		Antenna P.}, vol.~8, no.~12, pp. 887--893, Sept. 2014.
	
	\bibitem{YangAdams_SystematicShapeOptimizationOfSymmetricMIMOAntennasUsingCM}
	B.~Yang and J.~J. Adams, ``Systematic shape optimization of symmetric {MIMO}
	antennas using characteristic modes,'' \emph{IEEE Trans. Antennas Propag.},
	vol.~64, no.~7, pp. 2668--2678, July 2016.
	
	\bibitem{Krewski_2011_TCM_MIMO}
	A.~Krewski, W.~Schroeder, and K.~Solbach, ``Bandwidth limitations and optimum
	low-band {LTE MIMO} antenna placement in mobile terminals using modal
	analysis,'' in \emph{Proceedings of the 5th European Conference on Antennas
		and Propagation (EUCAP)}, Apr. 2011, pp. 142--146.
	
	\bibitem{Peitzmeier_Manteuffel-UpperBoundsForUncorrelatedPorts2019}
	\BIBentryALTinterwordspacing
	N.~Peitzmeier and D.~Manteuffel, ``Upper bounds and design guidelines for
	realizing uncorrelated ports on multimode antennas based on symmetry analysis
	of characteristic modes,'' \emph{IEEE Trans. Antennas Propag.}, vol.~67,
	no.~6, pp. 3902--3914, June 2019. [Online]. Available:
	\url{https://doi.org/10.1109/tap.2019.2905718}
	\BIBentrySTDinterwordspacing
	
	\bibitem{PapadimitriouSteiglitz_CombinatorialOptimization}
	\BIBentryALTinterwordspacing
	K.~S. Christos H.~Papadimitriou, \emph{Combinatorial Optimization}.\hskip 1em
	plus 0.5em minus 0.4em\relax Dover Publications Inc., 1998. [Online].
	Available:
	\url{https://www.ebook.de/de/product/3321118/christos_h_papadimitriou_kenneth_steiglitz_combinatorial_optimization.html}
	\BIBentrySTDinterwordspacing
	
	\bibitem{2005_ManteghiRahmatSamii_TARC}
	M.~Manteghi and Y.~Rahmat-Samii, ``Multiport characteristics of a wide-band
	cavity backed annular patch antenna for multipolarization operations,''
	\emph{IEEE Trans. Antennas Propag.}, vol.~53, no.~1, pp. 466--474, Jan. 2005.
	
	\bibitem{GustafssonTayliEhrenborgEtAl_AntennaCurrentOptimizationUsingMatlabAndCVX}
	\BIBentryALTinterwordspacing
	M.~Gustafsson, D.~Tayli, C.~Ehrenborg, M.~Cismasu, and S.~Norbedo, ``Antenna
	current optimization using {MATLAB} and {CVX},'' \emph{{FERMAT}}, vol.~15,
	no.~5, pp. 1--29, May--June 2016. [Online]. Available:
	\url{http://www.e-fermat.org/articles/gustafsson-art-2016-vol15-may-jun-005/}
	\BIBentrySTDinterwordspacing
	
	\bibitem{1978_Harrington_TAP}
	R.~Harrington, ``Reactively controlled directive arrays,'' \emph{IEEE Trans.
		Antennas Propag.}, vol.~26, no.~3, pp. 390--395, 1978.
	
	\bibitem{Harrington_FieldComputationByMoM}
	R.~F. Harrington, \emph{Field Computation by Moment Methods}.\hskip 1em plus
	0.5em minus 0.4em\relax Piscataway, New Jersey, United States: Wiley -- IEEE
	Press, 1993.
	
	\bibitem{Volakis_1998_FEM}
	J.~L. Volakis, A.~Chatterjee, and L.~C. Kempel, \emph{Finite Element Method
		Electromagnetics: Antennas, Microwave Circuits, and Scattering
		Applications}.\hskip 1em plus 0.5em minus 0.4em\relax Wiley -- IEEE Press,
	1998.
	
	\bibitem{Shilov_LinearAlgebra}
	G.~E. Shilov, \emph{Linear Algebra}.\hskip 1em plus 0.5em minus 0.4em\relax
	Dover, 1977.
	
	\bibitem{Wilkinson_AlgebraicEigenvalueProblem}
	J.~H. Wilkinson, \emph{The Algebraic Eigenvalue Problem}.\hskip 1em plus 0.5em
	minus 0.4em\relax Oxford University Press, 1988.
	
	\bibitem{CapekJelinek_OptimalCompositionOfModalCurrentsQ}
	M.~Capek and L.~Jelinek, ``Optimal composition of modal currents for minimal
	quality factor {Q},'' \emph{IEEE Trans. Antennas Propag.}, vol.~64, no.~12,
	pp. 5230--5242, Dec. 2016.
	
	\bibitem{Maseketal_ModalTrackingBasedOnGroupTheory}
	M.~Masek, M.~Capek, L.~Jelinek, and K.~Schab, ``Modal tracking based on group
	theory,'' \emph{IEEE Trans. Antennas Propag.}, vol.~68, no.~2, pp. 927--937,
	Feb. 2020.
	
	\bibitem{JelinekCapek_OptimalCurrentsOnArbitrarilyShapedSurfaces}
	L.~Jelinek and M.~Capek, ``Optimal currents on arbitrarily shaped surfaces,''
	\emph{IEEE Trans. Antennas Propag.}, vol.~65, no.~1, pp. 329--341, Jan. 2017.
	
	\bibitem{Balanis1989}
	C.~A. Balanis, \emph{Advanced Engineering Electromagnetics}.\hskip 1em plus
	0.5em minus 0.4em\relax Wiley, 1989.
	
	\bibitem{RosengrenKildal_RadiationEfficiencyCorrelationDiversityForMIMO}
	K.~Rosengren and P.-S. Kildal, ``Radiation efficiency, correlation, diversity
	gain and capacity of a six-monopole antenna array for a {MIMO} system:
	theory, simulation and measurement in reverberation chamber,'' \emph{{IEE}
		Proceedings - Microwaves, Antennas and Propagation}, vol. 152, no.~1, p.~7, 4
	2005.
	
	\bibitem{Balanis_Wiley_2005}
	C.~A. Balanis, \emph{Antenna Theory Analysis and Design}, 3rd~ed.\hskip 1em
	plus 0.5em minus 0.4em\relax Wiley, 2005.
	
	\bibitem{VolakisSertel_IntegralEquationMethodsForElectromagnetics}
	J.~L. Volakis and K.~Sertel, \emph{Integral Equation Methods for
		Electromagnetics}.\hskip 1em plus 0.5em minus 0.4em\relax Scitech Publishing
	Inc., 2012.
	
	\bibitem{SenoirVolakis_ApproximativeBoundaryConditionsInEM}
	T.~B.~A. Senior and J.~L. Volakis, \emph{Approximate Boundary Conditions in
		Electromagnetics}.\hskip 1em plus 0.5em minus 0.4em\relax IEE, 1995.
	
	\bibitem{Vandenbosch_ReactiveEnergiesImpedanceAndQFactorOfRadiatingStructures}
	G.~A.~E. Vandenbosch, ``Reactive energies, impedance, and {Q} factor of
	radiating structures,'' \emph{IEEE Trans. Antennas Propag.}, vol.~58, no.~4,
	pp. 1112--1127, Apr. 2010.
	
	\bibitem{CismasuGustafsson_FBWbySimpleFreuqSimulation}
	M.~Cismasu and M.~Gustafsson, ``Antenna bandwidth optimization with single
	frequency simulation,'' \emph{IEEE Trans. Antennas Propag.}, vol.~62, no.~3,
	pp. 1304--1311, 2014.
	
	\bibitem{CapekJelinekHazdra_OnTheFunctionalRelationBetweenQfactorAndFBW}
	M.~Capek, L.~Jelinek, and P.~Hazdra, ``On the functional relation between
	quality factor and fractional bandwidth,'' \emph{IEEE Trans. Antennas
		Propag.}, vol.~63, no.~6, pp. 2787--2790, June 2015.
	
	\bibitem{RaoWiltonGlisson_ElectromagneticScatteringBySurfacesOfArbitraryShape}
	S.~M. Rao, D.~R. Wilton, and A.~W. Glisson, ``Electromagnetic scattering by
	surfaces of arbitrary shape,'' \emph{IEEE Trans. Antennas Propag.}, vol.~30,
	no.~3, pp. 409--418, May 1982.
	
	\bibitem{Pozar_MicrowaveEngineering}
	D.~M. Pozar, \emph{Microwave Enginnering}, 4th~ed.\hskip 1em plus 0.5em minus
	0.4em\relax Wiley, 2011.
	
\end{thebibliography}
\end{document}